\documentclass[twocolappendix]{emulateapj}
\usepackage{graphics}
\usepackage{subfigure}
\usepackage{amsmath}
\usepackage{color}

\newcommand{\bmath}[1]{\mbox{\boldmath{$#1$}}}

\newcommand{\del}{{\bf \nabla}}
\newcommand{\pp}{p^\prime}

\newcommand{\cs}{c_{\rm s}}

\newcommand{\qsh}{q_{\rm sh}}

\begin{document}

\title{The Mass and Size Distribution of Planetesimals Formed by the Streaming Instability.
II. The Effect of the Radial Gas Pressure Gradient}

\author{Charles P. Abod\altaffilmark{1,2}, Jacob B. Simon\altaffilmark{3,2,4}, Rixin Li\altaffilmark{5}, Philip J. Armitage\altaffilmark{6,7,2}, \\ Andrew N. Youdin\altaffilmark{5}, Katherine A. Kretke\altaffilmark{4}}

\email{jbsimon.astro@gmail.com}

\begin{abstract}
The streaming instability concentrates solid particles in protoplanetary disks, leading to gravitational collapse into planetesimals. Despite its key role in producing particle clumping and determining critical length scales in the instability's linear regime, the influence of the disk's radial pressure gradient on planetesimal properties has not been examined in detail. Here, we use streaming instability simulations that include particle self-gravity to study how the planetesimal initial mass function depends on the radial pressure gradient. Fitting our results to a power-law, ${\rm d}N / {\rm d}M_p \propto M_p^{-p}$, we find a single value $p \approx 1.6$ describes simulations in which the pressure gradient varies by $\gtrsim 2$. An exponentially truncated power-law provides a significantly better fit, with a low mass slope of $p^\prime \approx 1.3$ that weakly depends on the pressure gradient. The characteristic truncation mass is found to be $\sim M_G = 4 \pi^5 G^2 \Sigma_p^3 / \Omega^4$. We exclude the cubic dependence of the characteristic mass with pressure gradient suggested by linear considerations, finding instead a linear scaling. These results strengthen the case for a streaming-derived initial mass function that depends at most weakly on the aerodynamic properties of the disk and participating solids. A simulation initialized with zero pressure gradient---which is {\em not} subject to the streaming instability---also yields a top-heavy mass function but with modest evidence for a different shape. We discuss the consistency of the theoretically predicted mass function with observations of Kuiper Belt planetesimals, and describe implications for models of early stage planet formation.
\end{abstract} 


\altaffiltext{1}{Department of Astrophysical and Planetary Sciences, University of Colorado, Boulder, CO 80309-0391}
\altaffiltext{2}{JILA, University of Colorado and NIST, 440 UCB, Boulder, CO 80309-0440}
\altaffiltext{3}{Department of Physics and Astronomy, Iowa State University, Ames, IA, 50010, USA}
\altaffiltext{4}{Department of Space Studies, Southwest Research Institute, Boulder, CO 80302}
\altaffiltext{5}{Department of Astronomy and Steward Observatory, University of Arizona, 933 North Cherry Avenue, Tucson, AZ 85721}
\altaffiltext{6}{Center for Computational Astrophysics, Flatiron Institute, NY 10010}
\altaffiltext{7}{Department of Physics and Astronomy, Stony Brook University, Stony Brook, NY 11794}


\section{Introduction} 

An outstanding issue in planet formation is how small particles of size $\sim$mm--cm grow to larger $\sim$1--100~km scale planetesimals. While growth through sticking is efficient for $\mu$m-mm scale particles \citep{blum18}, no viable path is known that leads to planetesimal formation via two-body collisions across all radii in the disk. In particular, silicate aggregates that may be representative of solids interior to the snow line fragment for mutual collision velocities $\Delta v \sim 1 \ {\rm m \ s^{-1}}$ \citep{guttler10}. For reasonable values of the disk turbulence parameter $\alpha$ and sound speed $c_s$ this threshold is substantially smaller than the peak collision speeds, $\Delta v \sim \alpha^{1/2}c_s$, predicted for a turbulent flow \citep{ormel07}. Icy materials exhibit higher threshold fragmentation velocities \citep{wada09,gundlach15}, and given enough time collisions could form larger bodies. In the presence of radial drift \citep{whipple72,weidenschilling77b}, however, even icy particle growth is limited to mm-cm sizes by the competition between inward drift and growth \citep{birnstiel12}.

A solution to these problems emerges when one considers not only the aerodynamic drag on the particles from the gas, but also the feedback of the particles onto the gas. The resulting equilibrium solution for radial drift in a vertically unstratified disk \citep{nakagawa86} is linearly unstable to the streaming instability \citep{youdin05}, which leads to particle clumping \citep{youdin07a,johansen07a,bai10c}. Under conditions that can plausibly, but not inevitably, be attained within protoplanetary disks, the clumping becomes strong enough \citep{carrera15,yang17} that self-gravity takes over and a fraction of the solid mass becomes bound objects. The most massive of these objects could have the size of large Solar System asteroids or Kuiper Belt objects \citep{johansen07a,johansen12,johansen15,simon16a,simon17,schafer17}.

Given the difficulty of directly observing planetesimals in the process of forming, one of the key routes to testing models is via their predictions for the initial mass or size distribution of planetesimals. The initial mass distribution is a direct input to models of the current size distribution of small bodies in the Solar System \citep{morbidelli09}, and it has an important, if indirect, impact on the predicted growth rates of giant planet cores \citep{pollack96,fortier13}, on the evolution of debris disks \citep{krivov18}, and on the abundance of interstellar planetesimals \citep{moromartin09,raymond18}.

Simple considerations suggest that the streaming-derived mass function might depend on as many as four independent parameters. The model linear problem \citep{youdin05}, which considers aerodynamically coupled dust and gas within an {\em unstratified} disk with zero intrinsic turbulence, yields growth rates and unstable modes that are functions of the dimensionless stopping time $\tau$, the solid-to-gas ratio $Z$ (more accurately the local solid-to-gas ratio, which depends on $Z$ as well as the particle scale height), and the radial pressure gradient parameter $\Pi$ (we defer to \S\ref{method} the formal definition of these parameters, but note here that $\Pi = \eta v_K / c_s$, where $\eta v_K$ is the difference between the gas velocity and the local Keplerian speed). The strength of intrinsic disk turbulence may also play a role, by controlling the thickness of settled particle layers \citep{dubrulle95}, acting as an effective diffusivity on small scales \citep{youdin05}, and strongly reducing the linear growth rate of the instability \citep{umurhan19}\footnote{ However, \cite{yang18} find that the presence of magnetically driven turbulence does not preclude strong clumping, despite reduced particle settling.}

The relationship between the instability of the unstratified model system (specified by $\tau$, $Z$, and $\Pi$) and the properties of planetesimals forming in real disks is unclear. Several additional pieces of physics need to be considered. The vertical component of stellar gravity leads to particle settling, and although this can be balanced by turbulent diffusion (leading to an equilibrium state) the linear theory including stratification has not been worked out. The combination of particle self-gravity (which is also not part of the model streaming problem) and aerodynamic drag leads to secular gravitational instability \citep{ward76,youdin11,takahashi14}, with distinct linear properties. Finally, gravitational collapse requires such high particle over-densities that expectations grounded in the linear physics may be a poor guide to planetesimal properties, even absent the extra physical effects discussed above. Indeed, numerical simulations suggest that some of the important linear parameters do not strongly modify the resulting mass function. Based on three high-resolution simulations that spanned a range of stopping time $6 \times 10^{-3} \leq \tau \leq 2$, \cite{simon17} found no evidence for significant variations with $\tau$. This led us to conjecture that the streaming-initiated planetesimal mass function might be universal, in the sense that some of its parameters (specifically, the low mass slope) are independent of the aerodynamic properties of the particles that participate in the instability.

Here, we extend our prior work to study how the planetesimal mass function depends on the radial pressure gradient in the disk. Since the linear scale of unstable modes in an unstratified disk without self-gravity is proportional to $\Pi$, there is a natural expectation that masses might scale as $\Pi^3$ \citep{youdin05,taki16}. 

In addition to testing this expectation, there are also observational motivations. Although $\Pi$ is a weak function of orbital radius for commonly adopted disk profiles, stronger variations are expected near the outer edge of the gas disk, which could correspond to the outer edge of the Kuiper Belt \citep{trujillo01}. Variations in $\Pi$ would also be expected if a significant fraction of planetesimals form from particles trapped within local pressure maxima \citep{whipple72,pinilla12,dittrich13}, caused for example by zonal flows \citep{johansen09a,simon14,bai14,bethune17}. Such traps may be responsible for the observed rings in a number of protoplanetary systems \cite[e.g.,][]{alma15,andrews16,isella16}.  If these rings are indeed sites for planetesimal formation, building a complete model of planetary system formation motivates us to understand the role of the radial pressure gradient in the birth of planetesimals.

This paper is organized as follows.  In \S\ref{method}, we describe our numerical algorithm in detail, define our initial conditions and choice of parameters, and describe the diagnostics we employ to characterize planetesimal formation. We present our results in \S\ref{results}. \S\ref{discussion} discusses the implications of those results, particularly in light of the observed size distribution of objects in the Kuiper Belt. We summarize and conclude in \S\ref{conclusions}. 


\begin{deluxetable*}{l|lc|ccccc}
\tabletypesize{\small}
\tablewidth{0pc}
\tablecaption{Streaming Instability Simulations\label{tbl:sims}}
\tablehead{
\colhead{Run}&
\colhead{$\Pi$\tablenotemark{a}}&
\colhead{$t_{\rm sg}$\tablenotemark{b}}&
\colhead{$p$\tablenotemark{c}}&
\colhead{$\pp$\tablenotemark{d}}&
\colhead{$M_0/M_G$\tablenotemark{e}}&
\colhead{$M_{50}/M_G$\tablenotemark{f}}&
\colhead{Mass Fraction}\\
\colhead{ }&
\colhead{ }&
\colhead{$(\Omega^{-1})$}&
\colhead{ }&
\colhead{ }&
\colhead{ }&
\colhead{ }&
\colhead{in Planetesimals} }
\startdata
P0-SG0 & 0 & 0 & 1.75 $\pm$ 0.03 & 1.52 $\pm$ 0.04 & 0.78 $\pm$ 0.14 & 0.57 & 0.81 \\
P0.0375-SG0 & 0.0375 & 0 & 1.64 $\pm$ 0.02 & 1.28 $\pm$ 0.03 & 0.26 $\pm$ 0.02 & 0.31 & 0.66\\
P0.05-SG0 & 0.05 & 0 & 1.60 $\pm$ 0.03 & 1.29 $\pm$ 0.03 & 0.35 $\pm$ 0.04 & 0.42 & 0.55\\
P0.0625-SG0 & 0.0625 & 0 & 1.59 $\pm$ 0.03 & 1.31 $\pm$ 0.04 & 0.32 $\pm$ 0.04 & 0.35 & 0.29\\
P0.075-SG0 & 0.075 & 0 & 1.58 $\pm$ 0.05 & 1.31 $\pm$ 0.07 & 0.56 $\pm$ 0.16 & 0.61 & 0.22\\
P0.0875-SG0 & 0.0875 & 0 & 1.58 $\pm$ 0.07& 1.33 $\pm$ 0.09 & 0.82 $\pm$ 0.27 & 1.0 & 0.12\\
P0.1-SG0-Lz0.4H & 0.1 & 0 & 1.49 $\pm$ 0.07 & 1.21 $\pm$ 0.07 & 0.44 $\pm$ 0.07 & 0.52 & 0.073\\
P0.0375-SG72 & 0.0375 & 72 & 1.56 $\pm$ 0.02 & 1.19 $\pm$ 0.02 & 0.16 $\pm$ 0.01 & 0.22 & 0.53 \\
P0.05-SG80 & 0.05 & 80 & 1.64 $\pm$ 0.03 & 1.30 $\pm$ 0.04 & 0.24 $\pm$ 0.04 & 0.26 & 0.28\\
P0.075-SG66 & 0.075 & 66 & 1.51 $\pm$ 0.06 & 1.21 $\pm$ 0.06 & 0.26 $\pm$ 0.05 & 0.38 & 0.062
\enddata
\tablenotetext{a}{\scriptsize Pressure gradient parameter, defined by Equation~\ref{headwind}}
\tablenotetext{b}{\scriptsize Time (in units of $\Omega^{-1}$) at which particle self-gravity is initiated}
\tablenotetext{c}{\scriptsize Best fit power law slope to simple power law model}
\tablenotetext{d}{\scriptsize Best fit power law slope to truncated power law model}
\tablenotetext{e}{\scriptsize Best fit truncation mass in truncated power law model}
\tablenotetext{f}{\scriptsize The mass at which $M_{p,{\rm tot, n}}(>M_p) = 0.5$, where $M_{p,{\rm tot, n}}(>M_p)$ is the total planetesimal mass greater than $M_p$ normalized to the total mass in planetesimals.}
\end{deluxetable*}


\section{Methods}
We describe our numerical methods and simulation diagnostics in detail below. For readers familiar with the field the setup is essentially identical to prior work \citep{simon16a,simon17}. We model a small but vertically stratified local volume of the disk, whose properties are specified by two parameters: the gas pressure gradient and a parameter describing the balance between self-gravity and shear. The solid component is specified by a single dimensionless stopping time and a ratio of the particle to gas surface densities. We solve the coupled system using {\sc Athena}, modeling the gas as a compressible isothermal fluid (ignoring magnetic fields and imposing no intrinsic disk turbulence) and the solids as an ensemble of individual particles. Particle self-gravity is implemented using a particle-mesh method with shearing radial boundary conditions. There are two key differences from our previous work \citep{simon16a,simon17}, both of which affect the analysis rather than the simulations themselves. We now identify gravitationally bound clumps of solids directly in three dimensions, rather than working with the projected surface density. We also apply a maximum likelihood estimator to fit our results to an exponentially truncated power-law, rather than the simple power-law that we restricted ourselves to before.

\label{method}

\subsection{Numerical Algorithm}

The simulations use the {\sc Athena} code to solve the equations of coupled particle motion
and hydrodynamics within the local, shearing box approximation. In this approximation we model 
a co-rotating disk patch whose size is small compared
to the radial distance $R_0$ from the central object.  This allows for 
the construction of a local Cartesian frame $(x,y,z)$, which
is defined in terms of disk's cylindrical coordinate system $(R,\phi,z^\prime)$ as $x=(R-R_0)$, $y=R_0 \phi$, and $z = z^\prime$.
This domain co-rotates around the central object at angular velocity $\Omega$, which corresponds
to the angular velocity at the center of the box (i.e., at $R_0$). A detailed
description of the shearing box can be found in \cite{hawley95a}. 

The equations of gas dynamics in this approximation are comprised of
the continuity and momentum equations:

\begin{equation}
\label{continuity_eqn}
\frac{\partial \rho}{\partial t} + \del \cdot (\rho {\bmath u}) = 0,
\end{equation}

\begin{eqnarray}
\label{momentum_eqn}
\frac{\partial \rho {\bmath u}}{\partial t} + \del \cdot \left(\rho {\bmath u}{\bmath u} + P {\bmath I} \right) 
& = & 2 \qsh \rho \Omega^2 {\bmath x} - \rho \Omega^2 {\bmath z} \nonumber \\
& & -2{\bmath \Omega} \times \rho {\bmath u} + \rho_p \frac{{\bmath v}-{\bmath u}}{t_{\rm stop}}
\end{eqnarray}

\noindent 
where $\rho$ is the mass density, $\rho {\bmath u}$ the momentum
density, $P$ the gas pressure, and ${\bmath I}$ is the identity matrix. The quantity $\qsh$ is
the shear parameter, defined as $\qsh = -d$ln$\Omega/d$ln$R$, which
we take to be $\qsh = 3/2$, as appropriate for a Keplerian disk.   From left to right, the source terms
in equation~(\ref{momentum_eqn}) correspond to radial tidal forces
(gravity and centrifugal), vertical gravity, the Coriolis force, and the feedback from the particle momentum onto the gas; 
$\rho_p$ is the mass density of particles, ${\bmath v}$ is their velocity, and $t_{\rm stop}$ is the timescale over which a particle will lose a factor of $e$ of its momentum due to gas drag.
 As we describe below, this feedback term is calculated at the location of every particle and then distributed to the gas grid points. 
 Finally, these gas dynamics equations are supplemented by an equation of state $P = \rho\cs^2$, where
$\cs$ is the isothermal sound speed. 

The left-hand side of the momentum equation (i.e., the motion of the gas in the absence of source terms)
is solved using a second-order accurate Godunov flux-conservative method, with the dimensionally unsplit corner transport upwind method
of \cite{colella90} and the third-order in space piecewise parabolic method of \cite{colella84}.
A detailed description of these components of {\sc Athena} along with various test problems
are given in \cite{gardiner05a}, \cite{gardiner08}, and \cite{stone08}.  There are several algorithmic additions employed to integrate these equations within the shearing box framework (thus handling the non-inertial terms), including orbital advection (the background Keplerian velocity is subtracted and integrated analytically; \citealt{masset00,johnson08}) and Crank-Nicholson differencing to preserve epicyclic energy to machine precision. Details for the implementation of these algorithms can be found in \cite{stone10}. 

Particles are treated in {\sc Athena} via the super-particle approach (i.e., each super-particle is a statistical representation of a number of smaller particles). The equation of motion for super-particle (hereafter, simply ``particle'' for simplicity) $i$ is given by

\begin{eqnarray}
\label{particle_motion}
\frac{d {\bmath v^\prime_i}}{dt} = 2\left( v^\prime_{iy} - \eta v_{\rm K}\right)& & \Omega \hat{\bmath x} - \left(2 - q\right) v^\prime_{ix} \Omega \hat{\bmath y} \nonumber \\ 
& & - \Omega^2 z \hat{\bmath z} - \frac{{\bmath v^\prime_i} - {\bmath u^\prime}}{t_{\rm stop}} + {\bmath F_{\rm g}}
\end{eqnarray}

\noindent
where the prime denotes a frame in which the background shear velocity has been subtracted, as part of the orbital advection scheme mentioned above. From left to right, the source terms are the radial acceleration of the particles due to the Coriolis effect, the gravitational + centrifugal force, and radial drift; azimuthal motion due to the Coriolis effect; vertical motion due to the central star's gravity; the gas drag; and the force due to particle self-gravity. The $\eta v_{\rm K}$ term accounts for the inward radial drift of particles resulting from a gas headwind, which in turn results from the radial pressure gradient.  Thus, $\eta$, which is a primary parameter of interest in this work, can also be described as the fraction of the Keplerian velocity by which the orbital velocity of particles is reduced (see Section~\ref{setup}).   To include the effects of this radial pressure gradient, we follow \cite{bai10a} and impose an inward force on the particles, resulting in the $\eta v_{\rm K}$ term as described above. This is equivalent to boosting our local domain into a frame moving slightly faster (by $\eta v_{\rm K}$) in the azimuthal direction than what would be present in a real disk, but the essential physics of differential motion between the gas and particles is accurately captured. 

Equation~(\ref{particle_motion}) is solved using the algorithms described in \cite{bai10a}. We integrate the equations of motion with a semi-implicit integration method combined with a triangular shaped cloud  (TSC) scheme to map the particle momentum feedback to the grid cell centers and inversely to interpolate the gas velocity to the particle locations (${\bmath u^\prime}$).  A more thorough description of the particle integration algorithm, as well as test problems, can be found in \cite{bai10a}. 

When particle self-gravity is activated, there is an additional force in equation~\ref{particle_motion}, ${\bmath F_{\rm g}}$, which is
found by solving Poisson's equation for particle self-gravity.  The details of
solving Poisson's equation can be found in \cite{simon16a}.  Briefly, we map the mass density of the particles to the grid cell centers using the TSC method and then calculate the gravitational potential by shifting the radial boundaries to be purely periodic (see \citealt{simon16a} and \citealt{hawley95a} for more details), applying a 3D FFT to calculate the gravitational potential in Fourier space, transforming the potential to real space using another 3D FFT, and then reversing the azimuthal shift to the original non-periodic frame.  In our simulations, the vertical boundaries are open, and as such, we apply a Green's function approach to the Poisson equation in the vertical direction \cite[see][]{koyoma09,simon16a}.  We then finite difference the potential to calculate the gravitational force at each grid cell center and then apply TSC to map these forces to the precise particle locations.  Tests of this algorithm are shown in \cite{simon16a}.

The boundary conditions for the gas and particles are shearing-periodic in the radial direction \citep{hawley95a}, purely periodic in the azimuthal direction, and a modified outflow condition in the vertical direction in which the gas density is extrapolated via an exponential function into the ghost zones \citep{simon11a,li18}.  This modified boundary condition prevents gas mass loss and spurious effects from developing near the vertical boundaries when the gas is in purely hydrostatic equilibrium.  Of course, once the system begins to evolve, this hydrostatic equilibrium will not be  maintained, and gas mass will be lost through the vertical boundaries \cite[see][]{li18}. To ensure mass conservation, at each time step, we renormalize the gas density in every cell so that the total gas mass is conserved.  The boundary conditions for the gravitational potential are essentially the same as the hydrodynamic variables; shearing-periodic in $x$ and purely periodic in $y$.  The vertical boundary conditions are open, and the potential is calculated in the ghost zones via a third order extrapolation.

\subsection{Initial Conditions, Parameters and Simulation Setup}
\label{setup}

\begin{figure}[ht!]
\begin{center}
\includegraphics[width=0.52\textwidth,angle=0]{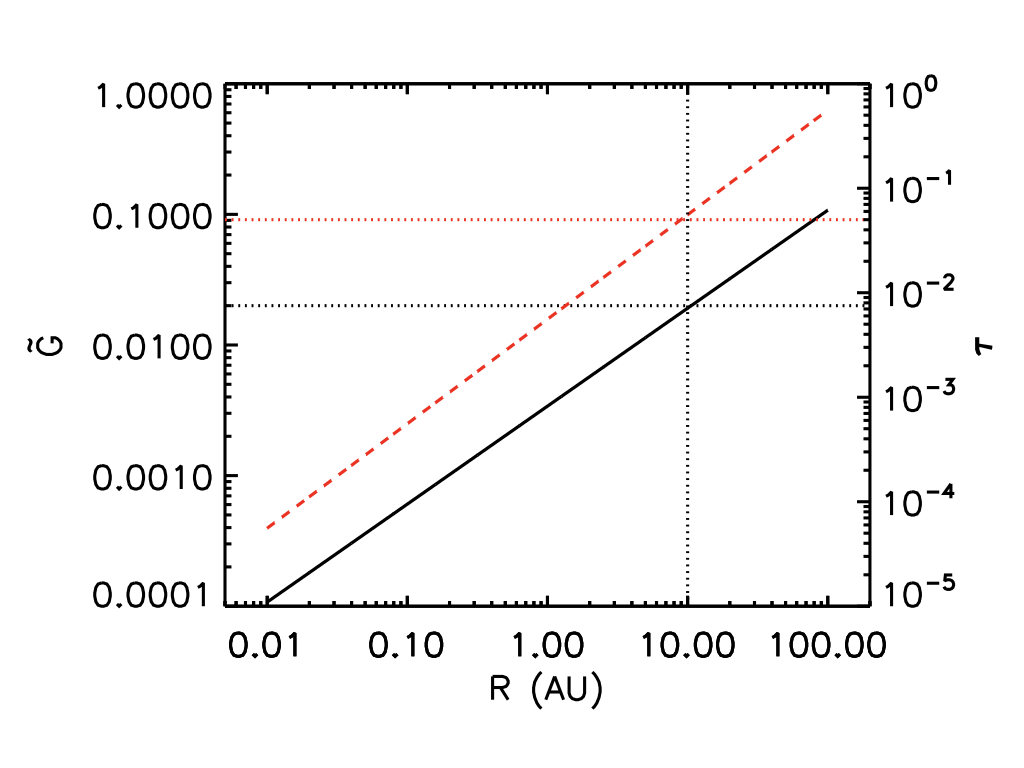}
\end{center}
\caption{Self-gravity parameter $\tilde{G}$ (solid, black curve) and dimensionless stopping time $\tau$ (dashed, red curve) as a function of radius (in AU) for a protoplanetary disk with surface density $\Sigma \propto r^{-1}$, disk mass 0.05$M_\sun$,  outer radius 500AU, and temperature profile $T = 280{\rm K} (r/{\rm AU})^{-1/2}$.  $\tau$ is calculated for the Epstein drag regime (this may not apply at small radii, but we are only interested in relatively large distances from the star) assuming a constant particle size of 2.5 mm. The higher (red) dotted line corresponds to the $\tau = 0.05$ value used in all of our simulations, and the lower (black) dotted line corresponds to $\tilde{G} = 0.02$ used in our simulations. The vertical
dotted line marks 10 AU, and corresponds to the radius at which $\tau = 0.05$ and $\tilde{G} = 0.02$.  At this radius, which is reasonable for planet formation to occur, our simulation parameters are physically realistic for typical protoplanetary disk conditions. \label{params}}
\label{beta}
\end{figure}

All of our simulations are initialized as follows.  The gas is in a hydrostatic (Gaussian) vertical profile, 

 \begin{equation}
 \label{gas_profile}
 \rho_g = \rho_0 \ {\rm exp}\left(\frac{-z^2}{2H^2}\right),
 \end{equation} 
 
 \noindent
 where $\rho_0$ is the mid-plane gas density. We choose code units so that the standard gas parameters are unity, $\rho_0 = H = \Omega = \cs = 1$.

Four dimensionless quantities characterize the evolution of the streaming instability and subsequent gravitationally collapse:  the
dimensionless stopping time,
\begin{equation}
 \tau \equiv t_{\rm stop} \Omega,
\end{equation}
which characterizes the aerodynamic interaction between a single particle species and the gas, the particle concentration,
\begin{equation}
 Z = \frac{\Sigma_p}{\Sigma_g},
\end{equation}  
which is the ratio of the particle mass surface density $\Sigma_p$ to the gas surface density $\Sigma_g$, and 
a radial pressure gradient parameter that accounts for the sub-Keplerian gas in real disks
\begin{equation}
\label{headwind}
\Pi \equiv \eta v_{\rm K}/\cs.
\end{equation}
This radial pressure gradient produces a headwind on the particles, which
has velocity a fraction $\eta$ of the Keplerian velocity.

There is an additional parameter that comes into play when self-gravity is included in the simulations: the relative
importance of self-gravity and tidal shear, which we quantify as 
\begin{equation}
\label{Gtilde}
\tilde{G} \equiv \frac{4\pi G\rho_0}{\Omega^2},
\end{equation}
Physically, varying $\tilde{G}$ corresponds to a change in gas density (and thus the particle density at constant $Z$)
and/or the strength of tidal stretching (i.e., changing $\Omega$). Another way to envision the importance of this parameter
is to consider how it would change with distance from the central star.  At larger radial distances, the tidal effects 
will be reduced compared to gravity, and thus $\tilde{G}$ will increase with radius.

In this work, we are primarily interested in the influence of $\Pi$ on the properties of planetesimals.  Thus, we fix the other parameters to be $\tau = 0.05$, $Z = 0.1$, and $\tilde{G} = 0.02$ (which corresponds to a Toomre $Q$ parameter for the gas $Q = 80$).  These particular choices are chosen to closely mimic reasonable disk parameters as much as possible.  In particular, Fig.~\ref{params} shows $\tilde{G}$ and $\tau$ (assuming 2.5mm sized particles, which is on the order of the maximum size particles can reach in a typical protoplanetary disk; \citealt{blum08,birnstiel11,zsom10}) for a typical protoplanetary disk with surface density $\Sigma \propto r^{-1}$ (assuming the disk has mass 0.05$M_\sun$ and an outer radius of 500 AU) and temperature profile $T = 280{\rm K} (r/{\rm AU})^{-1/2}$. As the figure shows, $\tilde{G} = 0.02$ near $R = 10$AU, and $\tau = 0.05$ corresponds to $\sim$ mm sized particles at this radius.  Thus, our values for these parameters are roughly consistent with that expected within the planet forming region given typical disk conditions.  In principle, a range of $\tau$ would be more appropriate since disk solids will not be confined to one size, 
 and indeed recent work suggests that this can appreciably change the dynamics \citep{bai10c,schaffer18,krapp19}. However, for the purposes of this work, in which we only vary $\Pi$, it is important to keep the other parameter choices as simple as possible. We thus defer the study of multiple particle species and its effect on the mass function to future work.  As shown by \cite{bai10b}, $Z$ must be larger than $\sim 0.05$ for $\Pi = 0.1$.  While these simulations were two-dimensional and included a range of $\tau$ values, we use them to further inform our parameter choices.  Thus, we choose the conservative $Z = 0.1$ for our concentration parameter. 

The particles are initially distributed uniformly in $x$ and $y$ with a Gaussian vertical profile. We choose the initial scale height of this vertical profile to be $H_p = 0.025H$.  We apply the Nakagawa--Sekiya--Hayashi (NSH) equilibrium solution in $x$ and $y$ \citep{nakagawa86} in order to ensure that the system starts in as close to an equilibrium configuration as possible.\footnote{Of course, in $z$, no such equilibrium exists.} To seed the streaming instability, we add random noise to the particle locations in all three dimensions.

One potential complication from these initial conditions, and in particular from using a factor of 10 larger concentration parameter relative to solar, which is $Z = 0.01$, is that once the particles sufficiently settle, the very thin, high density particle layer may overpower the Kelvin-Helmholtz instability that can in general prevent solids from settling \citep{cuzzi93,weidenschilling95}.  If this is the case, the particle layer will undergo gravitational collapse into planetesimals via the secular gravitational instability (hereafter, SGI; \citealt{ward76,youdin11,takahashi14}), which here refers simply to a regime of gravitational instability that is substantially modified by aerodynamic coupling to the gas.  

To circumvent this issue, our parameters, while physically motivated, were also chosen to make the growth rate of the streaming instability comparable to or larger than the settling rate (in the absence of turbulence) of $\tau = 0.05$ particles, which is $\approx 20 \Omega^{-1}$. At the given initial particle scale height, the ratio of particle mass density to gas mass density $\rho_p/\rho_g \approx 4$; extrapolating very roughly from Fig. 1 in \cite{youdin07a}, the growth time of the streaming instability is $\lesssim 10\Omega^{-1}$ on the scales which we can resolve.  This growth time is comparable to, and slightly less than the timescale over which the particles settle. Therefore, turbulence induced via the streaming instability should prevent the direct gravitational collapse of particles from the dense particle layer. 

In practice, however, we have found it difficult to guarantee that over-densities that lead to planetesimal formation result {\it only} from the streaming instability; in some of our simulations, the particle layer gets very thin before the streaming instability even sets in.\footnote{There are a number of simplifications in our estimates of the streaming growth rates, including the neglect of vertical gravity and a very approximate extrapolation from the linear analysis of \cite{youdin07a}.}   To test for this potential complication, we also run a simulation with $\Pi = 0$.  In such a simulation, the streaming instability vanishes, and any planetesimal formation must be from SGI.  Thus, by comparing the evolution and planetesimal properties in the $\Pi = 0$ run with our other simulations, we can determine what role SGI is playing in our simulations. 

One of the goals of this work is to determine what role, if any, the linear regime of the streaming instability plays in planetesimal formation.  As a comparison with the simulations described thus-far, we have also carried out three additional simulations in which self-gravity remained off until the streaming instability was fully non-linear.  We then chose a point during the saturated state of the instability to switch on self-gravity, after which we characterized the formation of planetesimals.

Most of our simulations are run in a domain size matching our previous work \citep{simon16a,simon17}, $L_x\times L_y\times L_z$ = $0.2H\times0.2H\times0.2H$, at a resolution of $N_x\times N_y\times N_z = 512\times512\times512$ with the number of particles $N_{\rm par} = 1.35\times10^8$. However, for $\Pi = 0.1$, we find that a large, asymmetric vertical outflow develops that carries away particle mass. We have determined that this effect, which only appears for $\Pi = 0.1$ (likely due to more vigorous turbulence in this case; see Section~\ref{efficiency}), does not significantly change the shape of the initial planetesimal mass functions.  However, because of the decreased particle mass, fewer planetesimals are formed.  Given these considerations, our simulation with $\Pi = 0.1$ is run in a taller domain with $L_x\times L_y\times L_z$ = $0.2H\times0.2H\times0.4H$, at a resolution of $N_x\times N_y\times N_z = 512\times512\times1024$, with the same number of particles as in the smaller domain runs; $N_{\rm par} = 1.35\times10^8$. This taller domain does not experience any asymmetric outflow or particle mass loss.

All of our simulations are listed in Table~\ref{tbl:sims}.  The naming convention for the simulation is P\#--SG\#, where P correspond to the $\Pi$ value, and SG is the time at which self-gravity is switched on. As mentioned above, for each simulation in which the self-gravity is not on from the beginning of the run, a simulation without self-gravity on is run with otherwise identical parameters until the streaming instability reaches a statistically saturated state.  We do not include these ``base'' simulations in the table. Finally, the tall box simulation described above follows the same naming convention, but with ``--Lz0.4H" appended.

 \subsection{Diagnostics}
 \label{diagnostics}

In what follows, we employ several diagnostics to fully analyze our simulations.  In this subsection, we first describe our new numerical algorithm for detecting and characterizing planetesimal properties.  We then present two statistical methods (based on maximum likelihood estimation) that we use to characterize the mass function of the produced planetesimals.  Finally, we wrap up with a brief description of our mass scaling and the fiducial times at which we analyze our simulations.

\subsubsection{{\sc PLAN}}
\label{plan_describe}
In order to quantify the planetesimal properties, we use a new clump-finding algorithm, PLanetesimal ANalyzer ({\sc Plan}).  Briefly, {\sc Plan} is based on the halo finder HOP \citep{eisenstein98} and is designed to find self-bound clumps by analyzing 3D particle output of {\sc Athena}.  With its hybrid OpenMP/MPI parallelization scheme written in C++, {\sc Plan} is efficient and scalable to handle billions of particles as well as multiple snapshots simultaneously. 

Following the standard HOP method, {\sc Plan} first computes a density for each particle with a user-defined smooth kernel using a memory-efficient Barnes-Hut tree \citep{barnes86}.  Particles with densities higher than a customized threshold,  $8\rho_0/\tilde{G}$, are selected and chained up towards densest neighbors to create groups.  {\sc Plan} then merges/decomposes those groups by examining boundaries, saddle points, and total kinematic and gravitational energies to construct a list of bound clumps.  An unbinding procedure is then performed to remove contamination particles (e.g., those that are passing by clumps and are not bound). {\sc Plan} then collects possibly unidentified member particles within each clump's Hill sphere.  Currently, {\sc Plan} does not further identify sub-structures within clumps because (i) most bound planetesimals in our simulations are highly-concentrated without subhalo-rich hierarchical structures as seen in cosmological simulations, (ii) most of their masses collapse into regions comparable/smaller than one hydrodynamic grid cell, invalidating further identification.  A more detailed description of {\sc Plan} is given in \cite{li19}. 

In practice, {\sc Plan} can calculate the mass of planetesimals much smaller than our previous clump-finding method \cite[e.g.][]{simon17}. However, an analysis of these small planetesimals has been carried out in a related work (see \citealt{li19}). Here we instead choose a minimum planetesimal mass of 1\% of the maximum planetesimal mass in all of our simulations. In this way, we are sampling the same relative range of masses for each $\Pi$, consistent with our previous papers, which by necessity only examined a maximum planetesimal mass range of approximately two decades.

\begin{figure*}[ht!]
\begin{center}
\includegraphics[width=1\textwidth,angle=0]{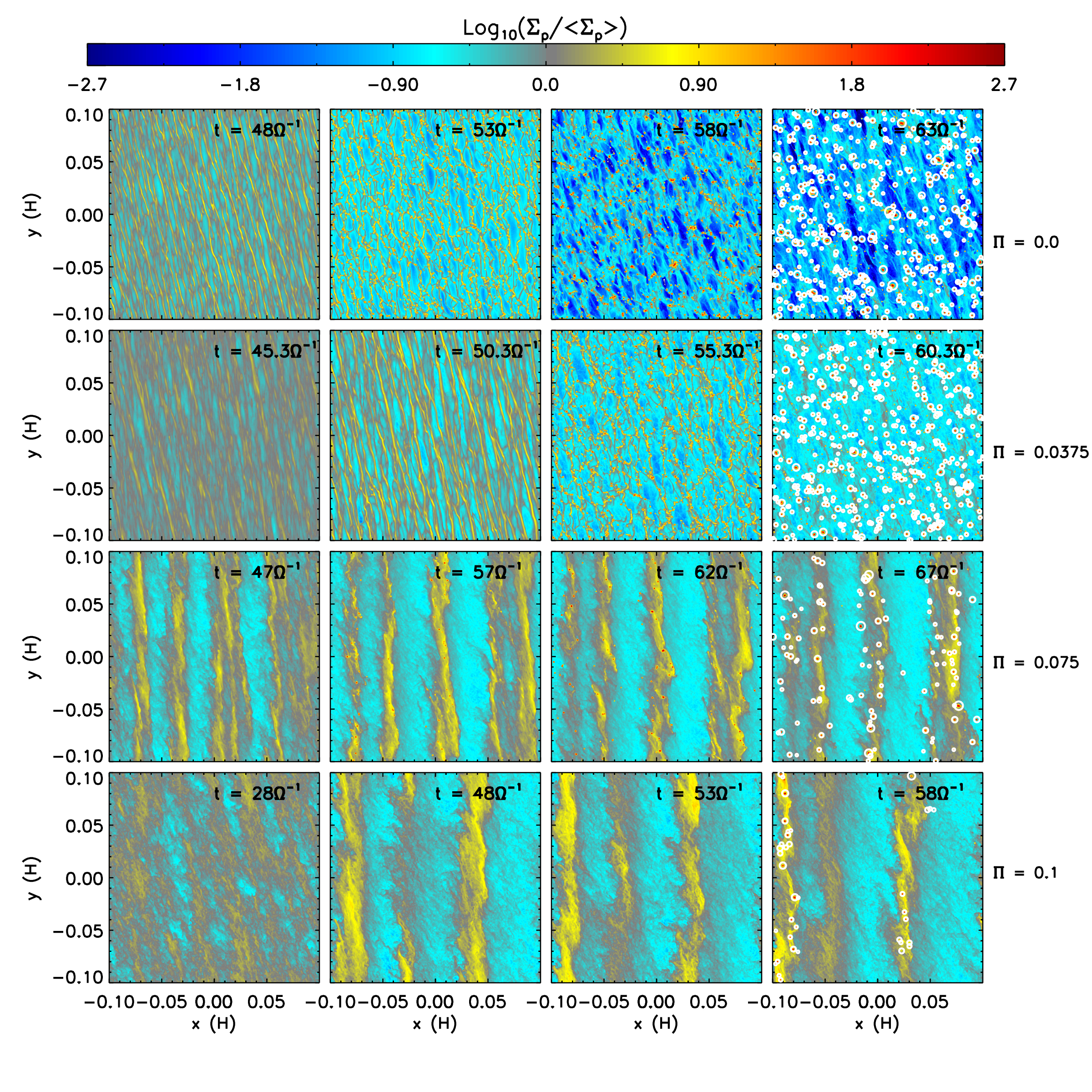}
\end{center}
\vspace{-0.4in}
\caption{Surface density of solids in the $x$--$y$ (orbital plane) for four of our SG0 simulations.  The color map is the logarithm of the surface density perturbations relative to the average surface density. Time increases from left to right, with the rightmost panel representing the fiducial time at which we analyze planetesimal properties. The top row corresponds to $\Pi = 0$, second row $\Pi = 0.0375$, third row $\Pi = 0.075$, and bottom row $\Pi = 0.1$.  
As $\Pi$ increases, the pre-collapse structure of solids moves towards larger scale, more axisymmetric structures, and the number of planetesimals decreases. In the rightmost column, white circles depict Hill spheres for a subset of planetesimals.}
\label{snapshots}
\end{figure*}

\subsubsection{Statistical Methods for Quantifying the Initial Mass Function}

In what follows, we use a maximum likelihood estimator (MLE) to determine the best fit parameters for two distinct functions.  The first is an {\it untruncated} power law function, hereafter generally referred to as the ``simple power law" (we use this to make contact with our previous results; \citealt{simon16a,simon17}), and the second
is a power law that is exponentially truncated (when written as the cumulative mass function).  In both cases, the MLE uses only the un-binned mass values to fit the relevant parameters: the power law indices and, in the case of the exponentially truncated mass function, a cut-off mass.

In constructing the mass functions, we first use a simple finite difference (after using {\sc Plan} to output the mass of the bound clumps) to determine an unsmoothed 
estimate of the differential mass function at masses corresponding to each 
planetesimal,

\begin{equation}
 \left. \frac{{\rm d}N}{{\rm d} M_{\rm p}} \right|_i = \frac{2}{M_{{\rm p}, i+1} - M_{{\rm p}, i-1}},
\end{equation} 

\noindent
as in \cite{simon16a} and \cite{simon17}.  The differential mass function can be written as

\begin{equation}
\label{mass_dist}
\frac{{\rm d}N}{{\rm d}M_p} = C_0 M_p^{-p},
\end{equation}

\noindent
where $C_0$ is a normalization constant set by the condition that the probability distribution function, $(1/n){\rm d}N/{\rm d}M_p$, where $n$ is the total number of planetesimals, integrated over all $M_p$ is unity. Thus,

\begin{equation}
\label{mle2_3}
    C_0 = n \left(p-1\right)M_{p,{\rm min}}^{p-1}
\end{equation}

\noindent
where $M_{p,{\rm min}}$ is the minimum value of the planetesimal mass in our data.
Following our previous two papers (\citealt{simon16a} and \citealt{simon17}), we use the maximum likelihood estimator (MLE) of \cite{clauset09} to determine $p$ in this simple untruncated power law function via

\begin{equation}
\label{mle_p}
p = 1 + n \left[\sum^n_{i=1} {\rm ln}\left(\frac{M_{p,i}}{M_{p,{\rm min}}}\right)\right]^{-1},
\end{equation}

\noindent
 The $1\sigma$
error in $p$ is

\begin{equation}
\label{mle_err}
\sigma_p = \frac{p-1}{\sqrt{n}}.
\end{equation}

\noindent
We next consider the exponentially truncated mass function, as written in the cumulative form (similar to that used in \citealt{johansen15} and \citealt{schafer17}), 

\begin{equation}
\label{cumulative}
    N(>M_p) = C_1 M_p^{-\pp+1} \exp \left(-M_p/M_0\right),
\end{equation}

\noindent
where $C_1$ is the normalization constant again set by the integrated probability equaling unity \citep{meerschaert12},

\begin{equation}
\label{mle2_3}
    C_1 = n M_{p,{\rm min}}^{\pp-1}e^{M_{p,{\rm min}}/M_0}, 
\end{equation}

\noindent
$\pp$ is the same power law index as above, but calculated via fitting to the truncated power law model, and $M_0$ is the characteristic mass for the exponential truncation. 

To calculate $\pp$, and $M_0$, we use another MLE approach, following the work by \cite{meerschaert12}.  In particular, from their equations 2.13--2.14, $\pp$ and $M_0$ can be found by solving the following set of nonlinear equations,

\begin{equation}
\label{mle2_1}
    \sum^{n}_{i=1}\frac{1}{\pp-1+M_{p,i}/M_0} = \sum^n_{i=1}{\rm ln}\left(\frac{M_{p,i}}{M_{p,{\rm min}}}\right),
\end{equation}

\begin{equation}
\label{mle2_2}
    \sum^{n}_{i=1}\frac{M_{p,i}}{\pp-1+M_{p,i}/M_0} = \sum^n_{i=1}\left(M_{p,i}-M_{p,{\rm min}}\right).
\end{equation}

In fitting the cumulative distributions below, we solve equations~\ref{mle2_1}--\ref{mle2_2} using the Newton-Rhapson method, after which we determine $C_1$ from equation~\ref{mle2_3}. To calculate the standard errors on these parameters, we use the simple bootstrap method of sampling with replacement, calculating the (assumed Gaussian) error as the standard deviation of the parameter samples produced by this method.

\subsubsection{Comparison of Fits}
\label{compare_fits}

To determine which fit is a better representation of our data, we use both the Bayesian Information Criterion (BIC) \citep{kass95}, 

\begin{equation}
\label{bic}
{\rm BIC} = K {\rm ln}\left(n\right) - 2 {\rm ln} \mathcal{L} , 
\end{equation}

\noindent
and the Akaike Information Criterion \citep{akaike74},

\begin{equation}
\label{aic}
{\rm AIC} =  2K - 2 {\rm ln} \mathcal{L}. 
\end{equation}

\noindent
In both criteria, $K$ is the number of parameters involved in the fit ($K = 1$ for the simple power law and $K = 2$ for the truncated power law), and ${\rm ln}(\mathcal{L})$ is the log-likelihood function, which is what is maximized in the MLE method (see, e.g.,  \citealt{meerschaert12,li19}.  ${\rm ln}(\mathcal{L})$ is calculated with the best fit parameters, and thus is in a sense a goodness-of-fit statistic (i.e., it gives the likelihood of the fit given the fitted parameters). Both criteria balance this goodness-of-fit with the complexity of the model, encapsulated by $K$. 

While the individual values of BIC and AIC cannot be interpreted, we calculate the relative difference between the simple power law and exponential power law as follows

\begin{equation}
\label{delta_bic}
\Delta {\rm BIC} = {\rm BIC}_{{\rm simple}} - {\rm BIC}_{{\rm exp}}
\end{equation}

\begin{equation}
\label{delta_aic}
\Delta {\rm AIC} = {\rm AIC}_{{\rm simple}} - {\rm AIC}_{{\rm exp}}
\end{equation}

\noindent
where the subscripts ``simple"  and ``exp" refers to the simple power law and exponentially truncated power law, respectively.  If $\Delta > 0$ (here referring to either or both $\Delta {\rm BIC}$ and  $\Delta {\rm AIC}$), the exponentially truncated power law is a better fit to the data under the criterion of interest and conversely, if  $\Delta  < 0$, the simple power law is a better fit.  See \cite{kass95} and \cite{akaike74} for more details. We quantify $\Delta {\rm BIC}$ and $\Delta {\rm AIC}$ in Section~\ref{truncated} below.

\subsubsection{Mass Scaling and Fiducial Times}
\label{mass_other}

In this work, we normalize all planetesimal masses to the dimensional mass for a self-gravitating particle disk. From balancing tidal and self-gravitational forces, one gets, from the standard Toomre dispersion relation, a critical unstable wavelength of

\begin{equation}
\label{lambdag}
    \lambda_G = \frac{4\pi^2G\Sigma_p}{\Omega^2}.
\end{equation}

\noindent For length-scales smaller than $\lambda_G$, gravity will overpower tidal shear.  Thus, we can define a ``gravitational mass" as patch of the particle disk with surface density $\Sigma_p$ and diameter $\lambda_G$; 

\begin{equation}
\label{grav_mass}
M_G \equiv \pi \left(\frac{\lambda_G}{2}\right)^2 \Sigma_p = 4 \pi^5 \frac{G^2\Sigma_p^3}{\Omega^4} = \frac{\sqrt{2}}{2}\pi^\frac{9}{2}Z^3\tilde{G}^2M_H
\end{equation}

\noindent
where $M_H = \rho_0 H^3$ is a dimensional reference mass.  

Finally, in what follows we will analyze planetesimal properties at specific {\it fiducial times}. These fiducial times were chosen ``by eye" at a time
in which bound clumps had recently formed but were not too connected to the large-scale structure from which they formed. We then verified that the power law slope did not change appreciably at different snaphsots (see below). For reference, the fiducial times for the SG0 simulations are $t = 63 \Omega^{-1}, 60.3 \Omega^{-1}, 64 \Omega^{-1}, 64.5 \Omega^{-1}, 67 \Omega^{-1}, 62 \Omega^{-1}$, and $ 58\Omega^{-1}$ for runs P0-SG0, P0.0375-SG0, P0.05-SG0, P0.0625-SG0, P0.075-SG0, P0.0875-SG0, and P0.1-SG0-Lz0.4H, respectively.  For the late gravity runs, these times are $84\Omega^{-1}$, $94\Omega^{-1}$, and $80\Omega^{-1}$ for P0.0375-SG72, P0.05-SG80, and PG0.075-SG66, respectively.

\begin{figure}[ht!]
\begin{center}
\includegraphics[width=0.48\textwidth,angle=0]{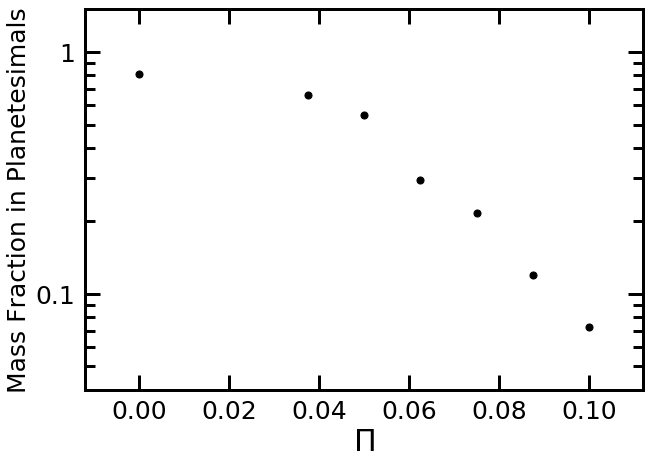}
\end{center}
\caption{Fraction of total solid mass within the domain converted to planetesimals versus $\Pi$ for all simulations with particle self-gravity always on.  The values plotted were taken at the fiducial times. The planetesimal formation efficiency strongly depends on $\Pi$, and for $\Pi = 0.1$, only 7.3\% of the solid mass is converted to planetesimals.}
\label{mass_frac_eta}
\end{figure}


\section{Results}
\label{results}

As we have done in our previous work \citep{simon16a,simon17}, our goal is to determine how planetesimal properties vary with the physical parameters relevant to the streaming instability.  Here, we focus on the pressure gradient parameter $\Pi$; we will first quantify the effects of a varying $\Pi$ on planetesimal formation and properties, followed by an examination, within the context of varying $\Pi$, of whether or not the time at which self-gravity is activated has any influence on the final properties of planetesimals.  We then discuss the special case of $\Pi = 0$, for which the streaming instability is absent. We finish the section with an independent verification of our results using a Kolmogorov-Smirnov test.

\subsection{Effect of the Pressure Gradient in the SG0 Runs}

Figure~\ref{snapshots} shows the particle surface density for four of our simulations: P0.0-SG0, P0.0375-SG0, P0.075-SG0, and P0.1-SG0-Lz0.4H.  In each case, we show the particle mass surface density in a series of snapshots, with time increasing from left to right.  The rightmost panel is the surface density at the fiducial times. As the figure shows, there is a dependence of the pre-collapse structure on $\Pi$.  First, for $\Pi \leq 0.0625$, the pre-collapse structure is largely non-axisymmetric, with small filaments that appear ``web-like" being azimuthally stretched (presumably by shear). As we describe further in Section~\ref{pi0} below, this non-axisymmetric structure is likely related to the effect of particle self-gravity.  For larger $\Pi$, the pre-collapse structures are mostly axisymmetric, and the width and separation of the structures increases with $\Pi$, a behavior broadly consistent with previous analyses \citep{li18,sekiya18}. 

Using {\sc Plan} to detect and analyze planetesimals in all of our simulations, we now quantify the properties of these planetesimals.

\begin{figure}[ht]
\begin{center}
\includegraphics[width=0.48\textwidth,angle=0]{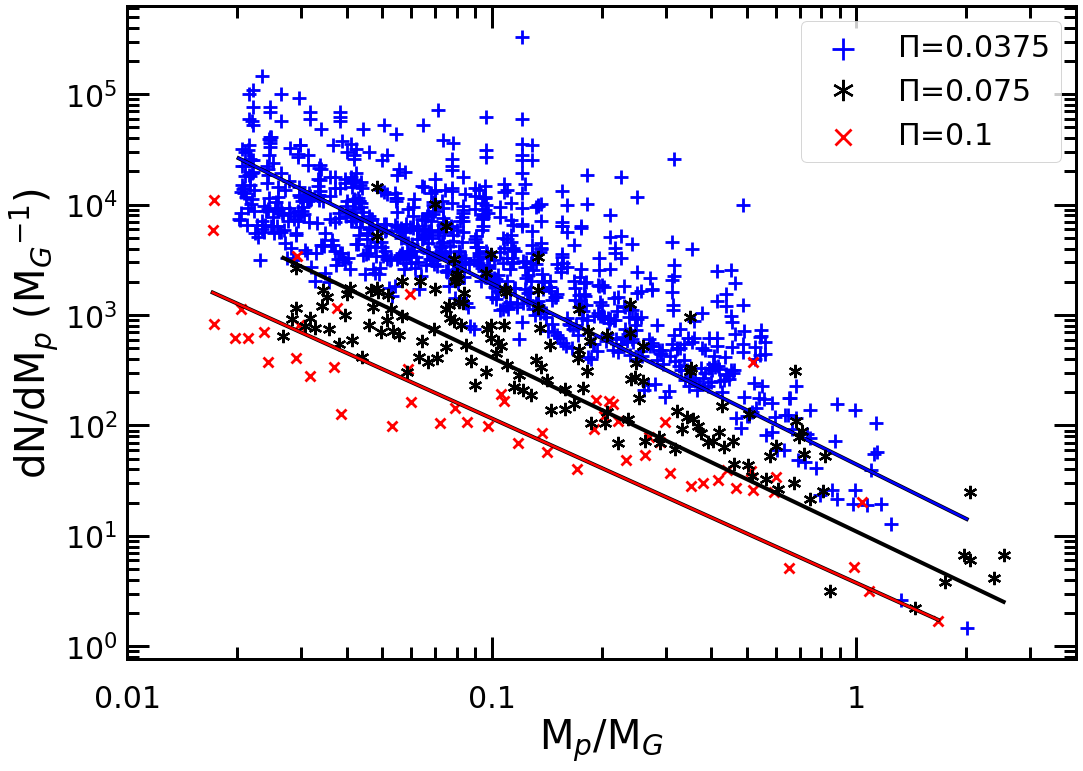}
\end{center}
\caption{Differential mass function for three of our simulations with $\Pi = 0.0375$ (blue crosses), $\Pi = 0.075$ (black asterisks), and $\Pi = 0.1$ (red x's). Over-plotted on each set of points is a line representing the best-fit. The masses are given in units of the gravitational mass (Equation~\ref{grav_mass}). As $\Pi$ increases, fewer planetesimals are produced, but the power law index is approximately constant.}
\label{dist_diff}
\end{figure}

\subsubsection{Planetesimal Formation Efficiency}
\label{efficiency}

From Fig.~\ref{snapshots} it is clear that the number of planetesimals formed decreases with increasing $\Pi$. This result can be further quantified by examining the fraction of solid mass converted to planetesimals (the planetesimal formation efficiency), as shown in Fig.~\ref{mass_frac_eta} and displayed in Table~\ref{tbl:sims}. There is a steep dependence of the planetesimal formation efficiency on $\Pi$, and for $\Pi = 0.0375$, 66\% of the mass has been converted into $\sim 10^3$ planetesimals.  At the other extreme, for $\Pi = 0.1$, only $\approx 7\%$ of solid mass is converted to $\sim 60$ planetesimals.\footnote{As described below, since we remove mass below 1\% of the maximum mass, these are lower limits to the total number of planetesimals. However, since the cumulative distribution begins to flatten in most cases, it is unlikely that the total number of planetesimals is significantly larger than this. Furthermore, we have verified that this mass cut-off does not strongly change the mass fraction converted to planetesimals.}

We have examined the turbulent kinetic energy of the gas for all of our simulations and have found that there is a clear increase with $\Pi$.  This result suggests that for larger $\Pi$, the more vigorous turbulence that is present may prevent a larger number of planetesimals from forming.  While there is some analytic and numerical work that is broadly consistent with this result (Armitage et al, in prep, Gole et al., in prep), a more detailed study of how turbulence affects planetesimal formation will be addressed in future work.

This strong dependence on planetesimal formation efficiency may have profound implications for the initial population of planetesimals in forming planetary systems, since $\Pi$ depends on radius, albeit weakly.  We should caution, however, that for $\Pi > 0.0625$, there remain significant axisymmetric structures after planetesimals have already formed.  It is entirely possible that planetesimal formation will continue with more planetesimals ultimately spawning from these structures, pushing the planetesimal formation efficiency higher for large $\Pi$.  Running these simulations further to test this hypothesis is very computationally expensive, and as such, we leave addressing this question to future work. 

\subsubsection{Initial Mass Function: Simple Power Law}

A key diagnostic in planetesimal formation studies is the distribution of masses, as it allows for a more direct comparison with Solar System planetesimal populations and can thus constrain planetesimal formation mechansisms. In Fig.~\ref{dist_diff}, we show the differential mass function for three of our simulations. Displaying the data in this way is done to make contact with our previous results. For each run, we over plot power law functions with the best-fit slope.  While the number of planetesimals differs drastically, as already discussed, the power law index remains approximately the same.  

\begin{figure}[ht!]
\begin{center}
\includegraphics[width=0.48\textwidth,angle=0]{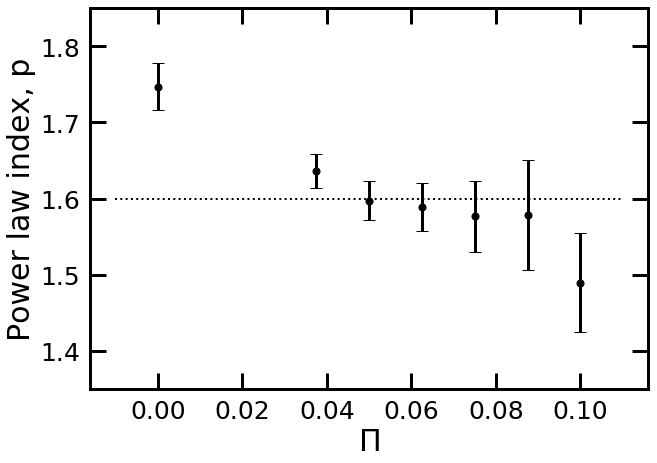}
\includegraphics[width=0.48\textwidth,angle=0]{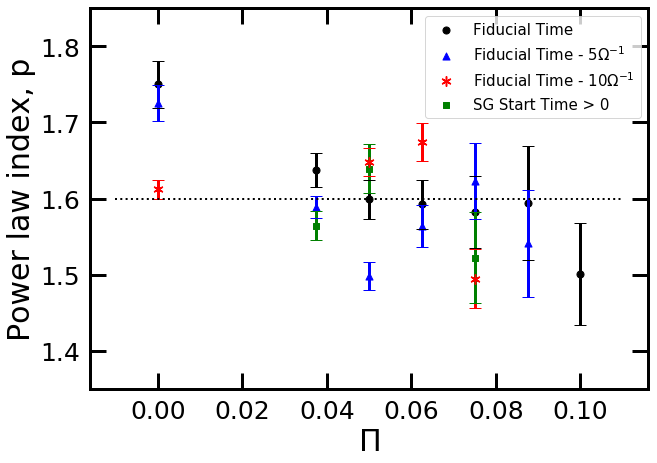}
\end{center}
\caption{Best fit power law index $p$ as a function of $\Pi$ for all of our simulations. The top panel shows $p$ at the fiducial times in the SG0 simulations.  The bottom panel combines these points with  $5\Omega^{-1}$ before the fiducial time (blue triangles),  $10\Omega^{-1}$ before the fiducial time (red asterisks), and simulations with self-gravity switched on at late times (green squares). Error bars denote $1\sigma$ uncertainty from equation~\ref{mle_err}.  Apart from $\Pi = 0$, all data points are consistent with $p = 1.5$--$1.6$ to within $1$--$2\sigma$. }
\label{p_eta}
\end{figure}

Figure~\ref{p_eta} further solidifies this result and displays $p$ for the entire range of $\Pi$ values explored here at the fiducial times and at  5$\Omega^{-1}$ and 10$\Omega^{-1}$ before the fiducial times (which corresponds to the third and second panels in Fig.~\ref{snapshots}, respectively).  The $p$ values at the fiducial times are also given in Table~\ref{tbl:sims}. With the exception of the $\Pi = 0$ simulation, $p = 1.5$--$1.6$ to within $2\sigma$.  This is in agreement with our previous results \citep{simon16a,simon17}; clearly $p \approx 1.6$ for a wide range of numerical and physical parameters.

\begin{figure}[ht!]
\begin{center}
\includegraphics[width=0.48\textwidth,angle=0]{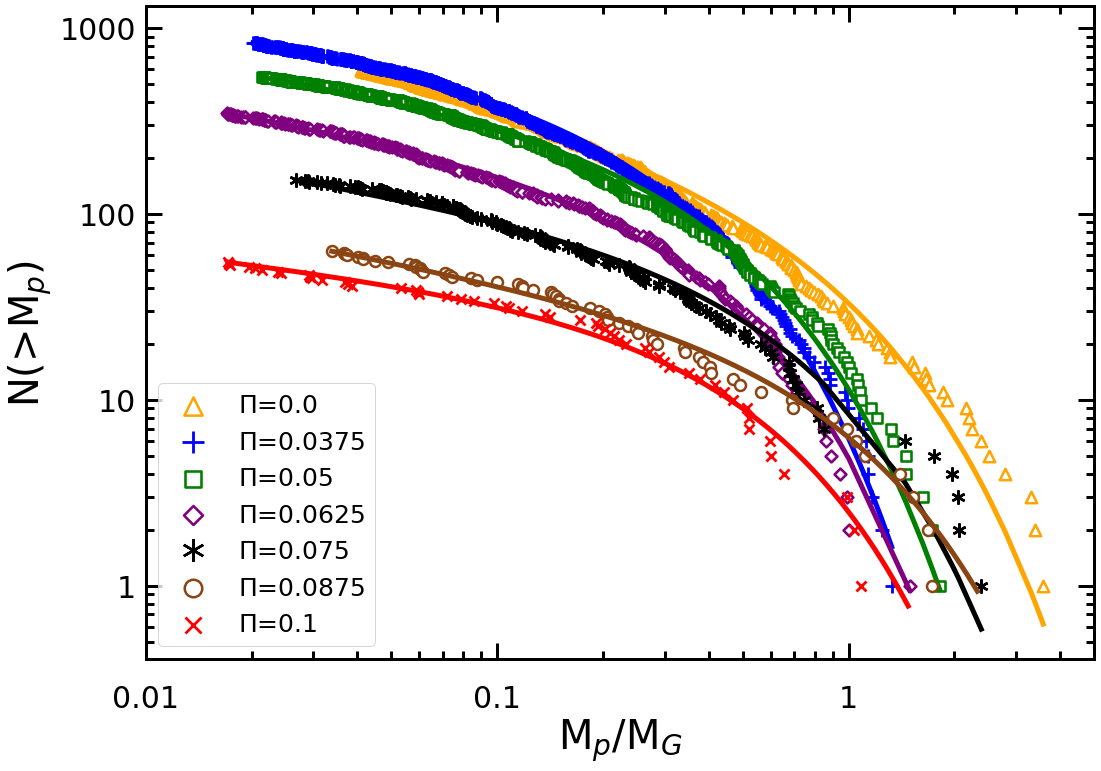}
\end{center}
\caption{Cumulative mass function for our simulations with $\Pi = 0.0$ (yellow triangles), 0.0375 (blue crosses), 0.05 (green squares), 0.0625 (purple diamonds), 0.075 (black asterisks), 0.0875 (brown circles), and 0.1 (red x's).  The masses are given in units of the gravitational mass (Equation~\ref{grav_mass}). Over plotted on each distribution is a best fit line of the same color (using equations~\ref{mle2_1}--\ref{mle2_3}) assuming an exponentially truncated power law. The exponentially truncated power law model provides a good fit to the data for $M_p < M_G$, though there remain moderate discrepancies at the highest masses.} 
\label{dist_cum}
\end{figure}

Another common diagnostic \citep{johansen15,simon16a,simon17,schafer17}, which removes most of the scatter inherent in the differential mass function, is the {\it cumulative} mass function. Figure~\ref{dist_cum} shows this cumulative mass function for all SG0 simulations.  Clearly, a simple power law does not fit the data over the entire range of masses when examining the cumulative mass function. Indeed, it appears as though $p$ itself depends on the planetesimal mass.  As we will show momentarily, these data can be better fit by an exponentially truncated power law.  However, we should first note that there may be different behavior below the mass cutoff described in Section~\ref{plan_describe}, as suggested by the ``flattening" of the cumulative mass function towards the smallest masses.  The issue of how the power law index may change at very small planetesimal masses is addressed in a related paper by our group \citep{li19}.  As such, we will not pursue this issue here. 

\subsubsection{Initial Mass Function: Truncated Power-law}
\label{truncated}

As is apparent from Fig.~\ref{dist_cum}, a single slope power law is too simplistic to accurately characterize the mass function over the entire range of masses probed. Here, we use equations~\ref{mle2_3}--\ref{mle2_2} to fit the mass function assuming an exponentially truncated power law in the cumulative distribution (i.e., equation~\ref{cumulative}).  The best fit curves are over-plotted in Fig.~\ref{dist_cum}. There is good agreement between this truncated power law model and the cumulative data for $M_p < M_G$, though we note that there remains moderate discrepancy at the high mass end of the distributions ($M_p \gtrsim M_G$), which could result from fewer planetesimals contributing to the fit in that region. 

\begin{figure}[ht!]
\begin{center}
\includegraphics[width=0.48\textwidth,angle=0]{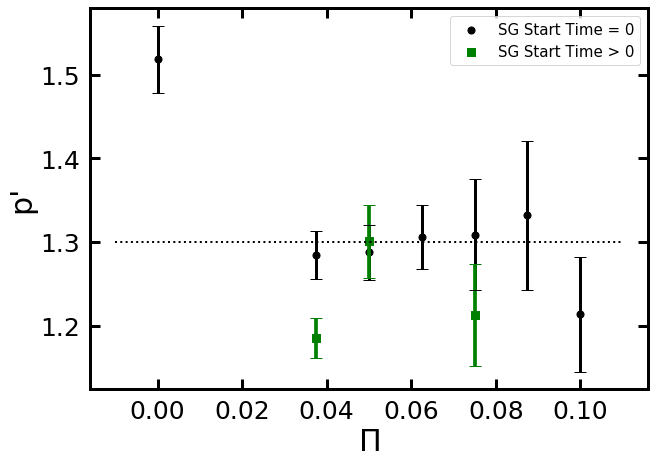}
\includegraphics[width=0.48\textwidth,angle=0]{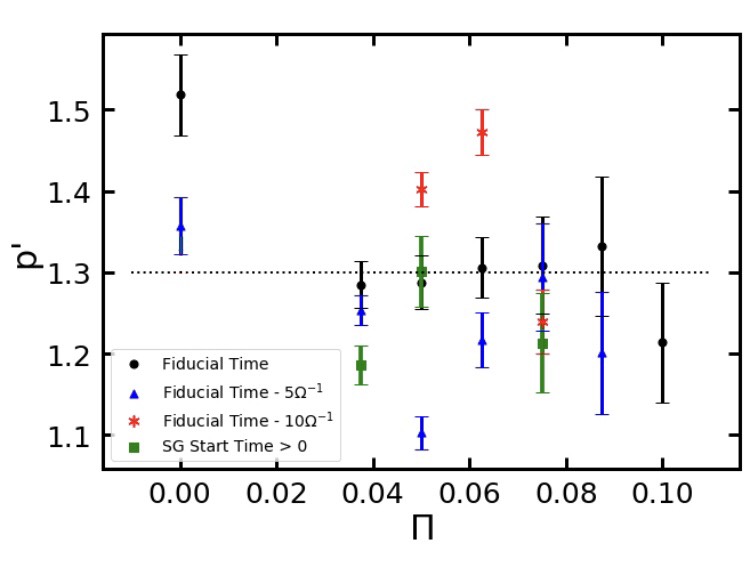}
\end{center}
\caption{Best fit power law index $\pp$, calculated from fitting to an exponentially truncated power law, as a function of $\Pi$ for all of the SG0 runs (black circles) and the late gravity runs (green squares).  The top panel shows only the fiducial times, whereas the bottom panel includes values from $5\Omega^{-1}$ (blue triangles) and $10\Omega^{-1}$ (red asterisks) before the fiducial times, respectively. Error bars denote $1\sigma$ uncertainty. Most data points are consistent with $p \approx 1.3$ to within 1--$2\sigma$. Exceptions include P0.0-SG0 (which is a special case) at the fiducial time; P0.0375-SG72, which displays a slightly lower $\pp$ value that is still consistent with $\pp = 1.3$ but to within $\approx 3\sigma$; and P0.0625-SG0 at 10$\Omega^{-1}$ before the fiducial time. 
 While $\pp$ is consistently less than its counterpart fitted to a simple power law ($p$) (see Fig.~\ref{p_eta}), $\pp$ depends only weakly on $\Pi$ for $\Pi > 0$ and is generally consistent with $\pp \approx 1.3$.}
\label{pp_eta}
\end{figure}

The exponentially truncated power law is fit by two parameters.  The first, $\pp$, is the index of the power law component, and is thus the differential power law slope {\it assuming the truncated power law model}. Fig.~\ref{pp_eta} shows $\pp$ (calculated at the fiducial times) versus $\Pi$ for all simulations, and these values are also displayed in Table~\ref{tbl:sims}. For $\Pi > 0$, $\pp \approx 1.3$ to within $1\sigma$ apart from $5\Omega^{-1}$ and 10$\Omega^{-1}$ before the fiducial time in P0.05-SG0 and 10$\Omega^{-1}$ before the fiducial time for P0.0625-SG0.  We are not entirely sure why these three points are outliers.  However, even in these instances, the values of $\pp$ do not vary by more than $\sim 20\%$.  Clearly, any dependence of $\pp$ on $\Pi$ is very weak.

 These shallower slopes, coupled with an exponential cut-off at a characteristic mass (which we discuss below), visually produces a better fit to the data than a simple power law.  To quantify this comparison, we use the BIC and AIC criteria described in Section~\ref{compare_fits} above. Table~\ref{tbl:comp_fits} displays the values of $\Delta$BIC and $\Delta$AIC for each value of $\Pi$. As these numbers clearly indicate, the exponentially truncated power law is a much stronger fit to the data than the simple power law, as both $\Delta {\rm BIC}$ and $\Delta{\rm AIC} \gg 1$ \citep{akaike74,kass95}.

\begin{deluxetable}{l|lcc}
\tabletypesize{\small}
\tablewidth{0pc}
\tablecaption{$\Delta$BIC and $\Delta$AIC Values \label{tbl:comp_fits}}
\tablehead{
\colhead{Run}&
\colhead{$\Pi$}&
\colhead{$\Delta$BIC}&
\colhead{$\Delta$AIC} }
\startdata
P0-SG0 & 0 & 63.4 & 67.8 \\
P0.0375-SG0 & 0.0375 & 237.7 & 242.4 \\
P0.05-SG0 & 0.05 & 151.2 & 155.5\\
P0.0625-SG0 & 0.0625 & 93.6 & 97.5 \\
P0.075-SG0 & 0.075 & 34.8 & 37.8 \\
P0.0875-SG0 & 0.0875 & 11.7 & 13.8\\
P0.1-SG0-Lz0.4H & 0.1 & 20.9 & 22.9
\enddata
\end{deluxetable}

In addition, the inclusion of the second parameter, which accounts for the steepening of the mass function towards higher masses, allows for a smaller power law index.  While the simple power law fit returned a value $p \approx 1.6$, which was largely determined by small planetesimals, the larger planetesimals inevitably play a role in making the simple power law fit steeper; this is no longer an issue as this steepening is fitted as the exponential tail in the truncated model. 

\subsubsection{Characteristic Planetesimal Masses}

The second parameter used in fitting the mass function to a truncated power law is $M_0$, the mass above which the mass function is substantially steepened. In fitting the data, $M_0$ is the only mass related parameter, and as such, it represents a characteristic mass for the distribution.  Furthermore, since the mass function is exponential for $M_p > M_0$, $M_0$ can be treated as a proxy for the maximum planetesimal mass. 
 
\begin{figure}[ht!]
\begin{center}
\includegraphics[width=0.48\textwidth,angle=0]{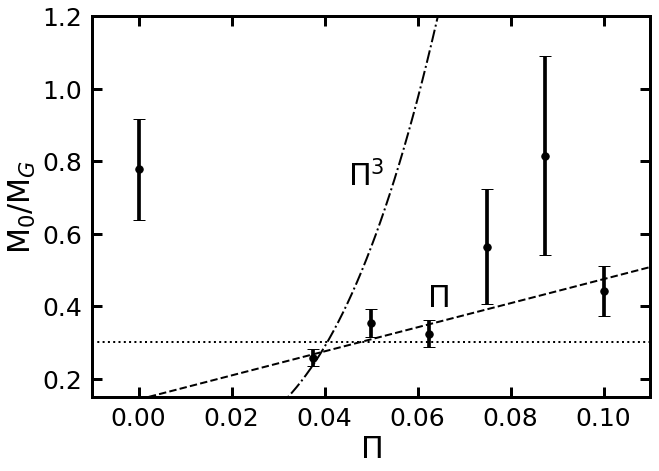}
\end{center}
\caption{Best fit characteristic planetesimal mass $M_0$ as a function of $\Pi$ for all SG0 simulations.  Error bars denote $1\sigma$ uncertainty. Apart from $\Pi = 0$, all data points are statistically consistent with either a constant $M_0 \approx 0.3M_G$ (dotted line) or a linear trend with $\Pi$ ($M_0 \propto \Pi$; dashed line). The data clearly rules out a $M_0 \propto \Pi^3$ dependence (dot-dashed line), as is expected if purely linear modes determined $M_0$.}
\label{m0_eta}
\end{figure}

Fig.~\ref{m0_eta} shows this characteristic mass (with associated error bars) as a function of $\Pi$. The masses (for $\Pi > 0$) lie within the range $M_0 \approx 0.25$--$0.3M_G$ to within $2\sigma$ and are thus potentially consistent with having no dependence on $\Pi$.  However, the data is also consistent with a linear increase of $M_0$ with $\Pi$ (dashed line in the Figure).
We calculated the $\chi^2$ values for both fits, finding a reduced $\chi^2 = 1.31$ and 3.19 for the linear and zero slope fits respectively.  Calculating the $F$ statistic, we find $F = 11.95$, which means that the linear model is a better representation of the data to within 97.5\% confidence (or between $2$--$3\sigma$). Thus, while we cannot completely rule out zero dependence of $M_0$ on $\Pi$, the data is significantly more consistent with a linear dependence.

 Similar conclusions follow from an analysis that does not depend on fitting a specific function to the data. Table~\ref{tbl:sims} shows the values of a quantity we label $M_{50}$, which is defined as the mass where,
\begin{equation}
 M_{p,{\rm tot,n}}(> M_p) = 0.5.   
\end{equation}
Here, $M_{p,\rm tot,n}(>M_p)$ is the total planetesimal mass greater than $M_p$ normalized to the total mass in planetesimals. We find that $M_{50}$ is on the order $M_G$ (ranging from 0.2--1$M_G$) in all simulations, and could be consistent with either a constant or relatively weakly increasing function of $\Pi$, though we have not quantified this statement since these values do not have errors.

As we have noted, there is no well-defined linear prediction for the unstable modes of the physical system that we are simulating (which is not in initial equilibrium, and which has self-gravity on from the start). However, to the extent that the system behaves similarly to the model linear problem (unstratified, and without self-gravity), we might expect the linear scales to be proportional to $\Pi$ \citep{youdin05}. This would result in a scaling of the characteristic planetesimal mass $M_p \propto \Pi^3$, as suggested by \cite{taki16}. Such a cubic scaling, normalized to the $\Pi = 0.0375$ result, is shown in Fig~\ref{m0_eta}. It is inconsistent with the available data.

Another way to see this result is to consider the ratio of the gravitational wavelength $\lambda_G$ (Equation~\ref{lambdag}) to the characteristic length-scale of the streaming instability $\eta r$,

\begin{equation}
\label{lambda_compare}
\frac{\lambda_G}{\eta r} = \sqrt{2\pi^3}\frac{\tilde{G}Z}{\Pi}.
\end{equation}

\noindent
If $\lambda_G \propto \eta r$ for all simulations, then the largest unstable mass (which is on the order of $M_0$) will scale with $\Pi^3$ (this is the essence of the \citealt{taki16} claim).  However, the values of $\lambda_G/\eta r$ range from 0.42 at $\Pi = 0.0375$ to 0.16 for $\Pi = 0.1$; $\lambda_G$ is clearly not proportional to $\eta r$, and the size scale of gravitationally bound clumps is not actually set by the value of $\Pi$ through the scales associated with the linear regime of the streaming instability.

However, it is worth noting that as $\Pi$ increases, so does the width of the pre-collapse filamentary structures (Fig.~\ref{snapshots}), suggesting that the linear increase in $M_0$ with
$\Pi$ could result from larger mass reservoirs feeding larger planetesimals.   Such a trend would be consistent with that observed by \cite{schafer17}.

In both of these characteristic masses, there is an uncertainty associated with continued planetesimal formation and continued accretion of solids onto and tidal stripping of already formed planetesimals. Longer duration simulations, possibly at higher resolution (to avoid unphysical merger events or accretion) are needed to measure any dependence of $M_0/M_G$ on $\Pi$ more robustly.  That being said, the scatter in the data should, to some extent, include this uncertainty, and even within the given scatter, a $\Pi^3$ dependence is ruled out. We conclude from this that the linear physics of the model streaming instability problem, specified by $\tau$, $\Pi$, and the local solid-to-gas ratio, does not simply translate into a prediction for the final sizes and masses of planetesimals formed in stratified self-gravitating simulations.
 
\subsection{Late Onset Gravity Simulations}
\label{nonlinear}

Simulations of planetesimal formation via the streaming instability start from simple initial conditions, and do not model processes such as particle growth and global radial drift that would bring the physical system to the point of instability. In particular, two types of simulations have been used in prior studies.
\begin{itemize}
    \item[(i)]Simulations in which particle self-gravity is turned on from $t=0$. The streaming instability (in its vertically stratified version) and SGI may both play a role in the formation of particle clumps and subsequent collapse. This choice is largely physically correct. However, by allowing collapse to occur rapidly---potentially before non-linear streaming-induced structures have fully formed---the results could be affected by the largely arbitrary initial conditions.
    \item[(ii)]Simulations in which the streaming instability is allowed to fully develop in the absence of self-gravity, which is only turned on at a later time. This choice allows the structure of two-phase turbulence in disks to fully develop prior to collapse (as likely occurs in reality), at the expense of excluding some of the physical effects of secular self-gravity. 
\end{itemize}
Since neither approach is fully realistic, comparing the two provides insight into whether the limitations of current simulations affect physical quantities of interest such as the mass function. In earlier lower resolution simulations we found that the slope of the mass function (fit as a simple power law) was not significantly altered by this numerical choice \citep{simon16a}. We revisit the issue here at high resolution in the context of models that vary $\Pi$.

\begin{figure*}[ht!]
\begin{center}
\includegraphics[width=0.48\textwidth,angle=0]{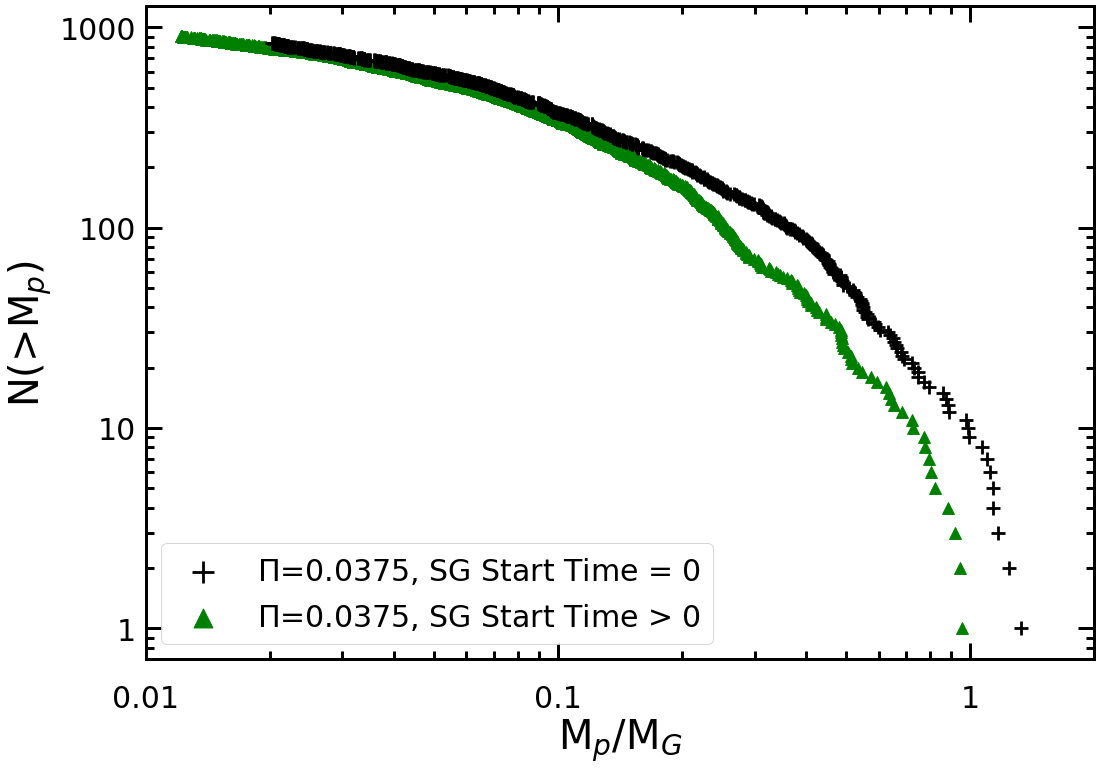}
\includegraphics[width=0.48\textwidth,angle=0]{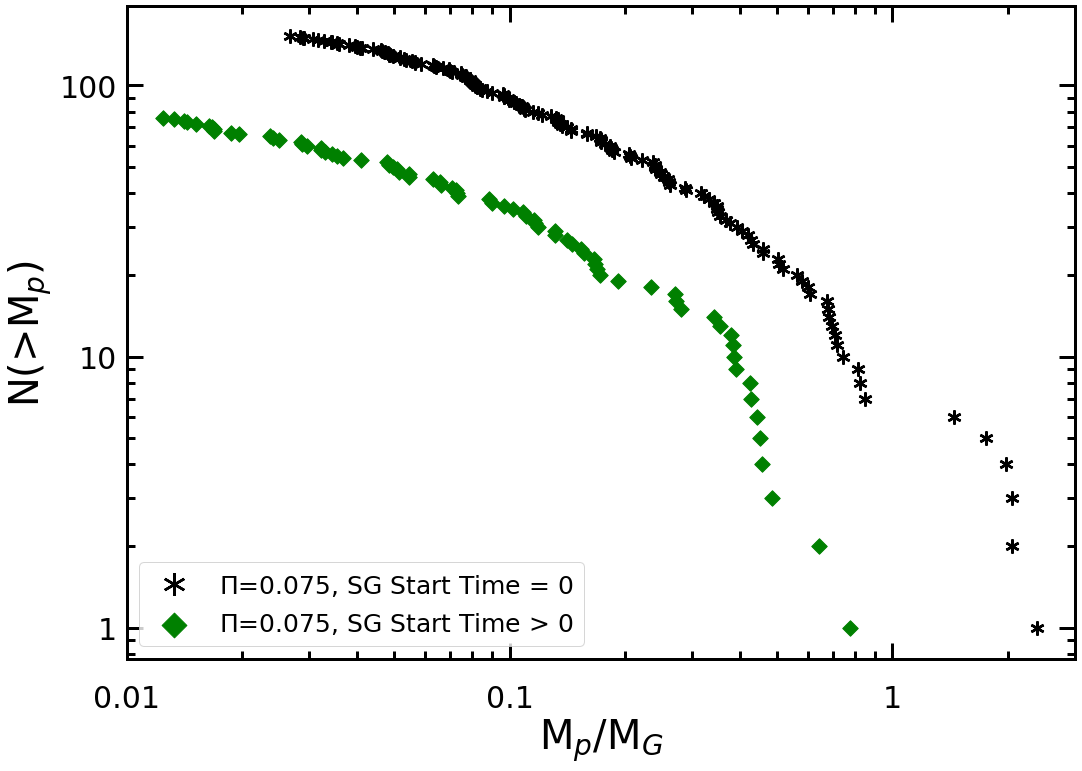}
\end{center}
\caption{Cumulative mass functions at the fiducial snapshots for simulations with $\Pi = 0.0375$ (left) and $\Pi = 0.075$ (right).  In the left (right) panel, the black crosses (asterisks) correspond to the SG0 simulations, whereas the green triangles (diamonds) correspond to the late gravity runs.  In both cases, the shape of the distributions are quite similar (particularly the slope at small masses), which suggests that the mass function is largely independent of the time at which self-gravity is activated.  However, there is evidence that the mass scale ($M_0$; equivalently where the mass function becomes exponentially steepened) is larger in the SG0 runs compared to the late-gravity runs, though given uncertainties associated with continued accretion and planetesimal formation, as discussd in the main text, such evidence remains tentative.}
\label{compare_gravity}
\end{figure*}

Our fiducial runs for this work all initiate self-gravity at $t=0$. We supplement these with three additional simulations in which we first let the streaming instability develop into its fully non-linear state without any particle self-gravity enabled. We then activated self-gravity at a time during the non-linear state for each simulation. In particular, for runs P0.0375-SG72, P0.05-SG80, and P0.075-SG66, we switched on self-gravity at $t = 72\Omega^{-1}, 80\Omega^{-1}$, and $66\Omega^{-1}$  (see Table~\ref{tbl:sims}), when $\rho_{p,{\rm max}}/\rho_0$ = 530, 540, and 193, respectively. Given the more than a factor of 10 enhancement of particle mass density over the initial values ($\rho_{p,{\rm max}}/\rho_0 \sim 10$), the system should be highly nonlinear at these times.  Furthermore, the streaming instability growth timescale with our chosen parameters is very roughly $\lesssim 10\Omega^{-1}$ (see Section~\ref{setup}). Thus, each of these simulations have evolved for many growth timescales before self-gravity is initiated, and we expect the system to be fully nonlinear when self-gravity is activated. 

Figure~\ref{compare_gravity} shows the cumulative mass functions at the fiducial times for two of these late gravity simulations compared with their SG0 counterparts. In both cases, the mass functions have similar slopes, though there is a difference in the mass values themselves, amounting to a factor of $\sim 3$ difference in P0.075-SG0 compared to P0.075-SG66. The corresponding parameters of the simple power law and truncated power law fits are plotted in 
Fig.~\ref{p_eta} and Fig.~\ref{pp_eta}.  As shown in these figures (and by examining the $p$ and $\pp$ values in Table~\ref{tbl:sims}), the power law slopes are consistent between the SG0 and late gravity runs to within $2\sigma$ for most of the runs and to within $\approx 3$--$4\sigma$ for $\Pi = 0.0375$. Moreover, when comparing the characteristic mass between the SG0 and late gravity simulations, $M_0$ is consistent with $0.2$--$0.3M_G$ to within $\approx 2\sigma$ with the exception of P0.0375-SG72, which is consistent with $0.2M_G$ to within $4\sigma$(see Table~\ref{tbl:sims}). We should note however, that in every case, $M_0$ in the late gravity simulations is less than $M_0$ in the corresponding SG0 run, a feature which is seen in Fig.~\ref{compare_gravity} as well. Nevertheless, given the error bars, the discrepancy between the model and data at masses greater than $M_G$, and the potential role of continued accretion (and possibly tidal interactions) as described above, there is, at best, a weak dependence of $M_0$ on the self-gravity start time. 

Our conclusion is that the interplay between the initial conditions and the time at which self-gravity is initiated do not significantly change the properties of the mass function.  This is consistent with our previous result \citep{simon16a}, derived from substantially lower resolution runs, that $p$ is largely independent of initial conditions and whether or not self-gravity is activated from the beginning of the run. Physically, it may suggest that the clumping of solids due to the non-self-gravitating streaming instability alone, and that caused by the combined action of streaming and secular self-gravity, are similar. More in-depth work, however, will be needed to determine precisely how the non-linear turbulence driven by the streaming instability in the presence of self-gravity leads to the planetesimal masses and mass functions that we see. 

\subsection{The Special Case of Zero Pressure Gradient}
\label{pi0}

As mentioned previously, we have carried out one simulation with $\Pi = 0$. Since, by definition, there is no gas-particle drift for $\Pi = 0$, the streaming instability is necessarily absent. The effects of SGI are however present. A direct comparison between the planetesimals that form in the $\Pi=0$ case and those that form for $\Pi > 0$ thus tests to what extent the streaming instability is present for the $\Pi > 0$ simulations versus SGI.

\begin{deluxetable*}{l|cccccc}
\tabletypesize{\small}
\tablewidth{0pc}
\tablecaption{Kolmogorov-Smirnov probabilities when comparing mass distributions between runs.\label{tbl:ks}}
\tablehead{
\colhead{ }&
\colhead{P0.0375-SG0}&
\colhead{P0.05-SG0}&
\colhead{P0.0625-SG0}&
\colhead{P0.075-SG0}&
\colhead{P0.875-SG0}&
\colhead{P0.1-SG0-Lz0.4H} }
\startdata
P0-SG0 & $4.1\times10^{-6}$ &  $7.9\times10^{-7}$ & $10^{-5}$ & $6.1\times10^{-5}$ & 0.007 & $10^{-5}$ \\
P0.0375-SG0 & x & 0.29 & 0.03 & 0.26 & 0.52 & $6\times10^{-4}$ \\
P0.05-SG0 & x & x & 0.12 & 0.59 & 0.84 & 0.002 \\
P0.0625-SG0 & x & x & x & 0.65 & 0.81 & 0.05 \\
P0.075-SG0 & x & x & x & x & 0.69 & 0.05\\
P0.0875-SG0 & x & x & x & x & x & 0.13
\enddata
\end{deluxetable*}

The top row of Fig.~\ref{snapshots} shows the particle mass surface density for $\Pi = 0$ for a series of time snapshots. There is considerable small scale structure that develops before planetesimals form, with many narrow filaments that appear elongated in azimuth (most likely due to shear). This mechanism generates a large number of planetesimals that are uniformly distributed throughout the computational domain.  The pre-collapse structure seen in this simulation is quite similar to that in P0.0375-SG0 (as well as in P0.05-SG0 and P0.0625-SG0; not shown in the figure). 

On the other hand, for the late-gravity runs, P0.0375-SG72 and P0.05-SG80, the particle surface density at early times (before self-gravity is activated) resembles the larger $\Pi$ simulations in that there are largely axisymmetric structures (though of significantly smaller radial scale compared to the larger $\Pi$ runs) induced that do not have the same ``web-like" shape demonstrated by the SG0 runs.  Evidently, self-gravity plays a role in determining the pre-collapse structure of the particles for $\Pi \leq 0.0625$. 

This result suggests that the mass function in the SG0 runs with $\Pi \leq 0.0625$ may be influenced by the same physics that determines the $\Pi=0$ mass function.  Indeed it may very well play some sub-dominant role in P0.0375-SG0; the value of $p$ is slightly larger than the other $\Pi > 0$ runs and the late gravity $\Pi = 0.0375$ simulation. For $\Pi > 0.0375$, we find $p$ is more than $\sim 3$--$4\sigma$ lower than the $\Pi = 0$ case, in which $p = 1.75 \pm 0.03$.   $M_0$ is also larger than the mass ($\sim 0.3M_G$) with which the $\Pi > 0$ simulations are statistically consistent. However, apart from $\Pi = 0.0375$, the $\Pi = 0$ $M_0$ is also statistically in agreement with the larger $\Pi$ simulations to within $2\sigma$.  Furthermore, the spread in $\pp$ values in all simulations, including that of $\Pi = 0$ is relatively large, suggesting that $\pp$ is not significantly different for $\Pi = 0$ compared to $\Pi > 0.$

Overall, there are indications that the mass function of $\Pi = 0$ is different than for $\Pi > 0$, but such evidence for this difference remains modest, given the spread in $\pp$ values.  At the current time, we do not fully understand why the values of $\pp$ have as large a spread as they do.  However, despite these differences, it is noteworthy that $p < 2$ for all $\Pi$ values.  That is, both the streaming instability simulations and the $\Pi=0$ run where streaming is absent produce a top-heavy mass function where most of the mass is in the largest planetesimals. Many of the qualitative implications discussed in the context of planetesimals forming via the streaming instability carry over directly to the case with zero pressure gradient.


\subsection{Robustness of Derived Fits}

For the results presented above we have fit the different models to planetesimal data that spans two orders of magnitude in mass. This choice was made to ensure uniformity across runs, and because we are confident that the least massive planetesimals remain well-resolved. We have briefly explored how the results change if we lower the low mass cut-off to less than the fiducial 1\%. Doing so we found modestly lower $\pp$ values (measured at the fiducial times), with $\Pi = 0$ dropping the most but remaining distinct from the $\Pi > 0$ runs. However, we also found that some of the fits to a truncated power law are not nearly as good (as judged by eye) when lowering the cut-off, making our statements about lower $\pp$ more tentative.

We do not have any strong theoretical reason to expect that the true mass function is a truncated exponential---we have adopted this function because it appears to be a reasonable approximation that can be described by only two parameters. More complex functions would yield better fits, and might justify the introduction of extra parameters. For example, exponentially truncated power laws with an additional exponent parameter (${\rm exp}[-(M_p/M_0)^\gamma]$) have been fit to planetesimal data by \cite{johansen15} and \cite{schafer17}. \cite{schafer17} find that an exponentially truncated power law with $\gamma \approx 0.3$--$0.4$ fit their data reasonably well, though they were unable to strongly constrain the low mass end of the mass function using their data.\footnote{It is worth noting, however, that for the same domain size and $\Pi$ value (and despite differences in $\tau$, $Z$, and numerical resolution), the characteristic mass in \cite{schafer17} (see ``run\_0.2\_640" in their Fig. 5 or Table 2) is consistent with $M_0$ in both P0.05-SG0 and P0.05-SG80, to within factors of order unity.} Further work is thus needed to determine what is the best description of the entire mass function forming from the streaming instability.

Given the above uncertainties the precise extrapolation of the derived mass function to lower masses (determined by the values of $p$ and $\pp$) remains unclear. However, we consider that the {\em consistency} of the derived parameters across different $\Pi > 0$ runs (i.e. $p$ for the simple power law and $\pp$ for the truncated model) to be more robust and physically significant. Both the fitting exercise described above, and the model-independent tests presented in \S\ref{ks}, lead to two conclusions. First, for the ``streaming unstable" runs with $\Pi > 0$, very similar mass functions are derived for all of the relevant streaming parameters studied to date \citep{simon16a,simon17}. We have interpreted this previously as pointing to the primary role played by nonlinear turbulence in determining the initial mass function of planetesimals. Second, the $\Pi=0$ run that is not unstable to streaming has a different mass function, though the differences here remain tentative.

\subsection{Model-Independent Verification of Results}
\label{ks}

As discussed above, the truncated power law fit (which is already significantly better than the simple power law fit) to the mass functions shows the largest discrepancy for $M_p \gtrsim M_G$; the fit nearly always leads to fewer planetesimal numbers at the highest masses. To determine how such discrepancies might affect our results, we present here a model-independent test.  Namely, we determine the probability that any two mass functions 
are derived from the same distribution via a Kolmogorov-Smirnov (KS) test, in which we normalize the masses by the maximum mass for each simulation and then apply the KS test in logarithmic space. 

Table~\ref{tbl:ks} displays the probability value from the KS test for every combination of pairs between the SG0 simulations.  While a large range of probabilities exist for the runs with $\Pi > 0$, most of the probabilities are $> 0.1$.  In contrast, the KS probability from comparing P0-SG0 to the other simulations are mostly $<10^{-4}$ (with the exception of 0.007 for P0.0875-SG0 compared to P0-SG0). This large discrepancy supports our claim that the mass function for $\Pi = 0$ is different than for $\Pi > 0$.  

It is also worth noting that the $\Pi = 0.1$ run shows both moderate and low probabilities, ranging from $6\times10^{-4}$ to 0.13, when compared against other $\Pi > 0$ runs.  This is consistent with our general finding that both $p$ and $\pp$ are lower for $\Pi = 0.1$ compared to the other $\Pi > 0$ simulations. However, given the large error bars on the fit parameters and the range in probability values from the KS test, we report that there is, at best, tentative evidence for $\Pi = 0.1$ producing a different mass function than the other runs with $\Pi > 0$.  Regardless of the robustness of such evidence, the deviation of $\Pi = 0.1$ from the other runs appears to be modest.


\section{Compatibility with observations}
\label{discussion}
The most direct constraints on the initial planetesimal mass function come from observations of populations of small bodies---asteroids and Trans-Neptunian Objects (TNOs)---in the Solar System. Suppose that we fit (allowing for a varying power law index) the differential number distributions of these objects as power-laws in mass $M_p$, diameter $D$, and absolute magnitude $H$ (taking the $H$ band as a relevant band for TNOs),
\begin{eqnarray}
 \frac{{\rm d}N}{{\rm d}M_p} & \propto & M_p^{-p}, \\
 \frac{{\rm d}N}{{\rm d}D} & \propto & D^{-q}, \\
 \frac{{\rm d}N}{{\rm d}H} & \propto & 10^{\alpha H}.
\end{eqnarray} 
Then, if we assume constant albedo and constant density, there are simple relations linking the slope of the mass distribution $p$ to the slope of the size distribution $q$ and to the observationally accessible distribution of absolute magnitudes $\alpha$,
\begin{eqnarray}
 q & = & 3p - 2, \\
 q & = & 5 \alpha + 1.
\end{eqnarray} 
Fitting the streaming instability results to a simple power-law yields $p \simeq 1.6$, which implies that $q \simeq 2.8$ and that $\alpha \simeq 0.36$ \citep{johansen12,simon16a,schafer17}. The low mass end of the truncated power-law fit yields $p^\prime \simeq 1.3$, giving $q \simeq 1.9$ and $\alpha \simeq 0.18$.

Assessing the consistency of these predictions with observations is not an easy task. The smallest bodies, especially in the asteroid belt, are collisionally evolved \citep{dohnanyi69}, while the largest bodies may have grown via accretion of planetesimals or pebbles \citep{ormel10,johansen15}. Both of these modifiers depend on the initial mass in small bodies in the belts, which is a large but uncertain multiple of the current mass.

\subsection{Asteroid Belt size distribution}
The degree to which the current asteroid belt size distribution is consistent with streaming predictions has been studied by \cite{morbidelli09} and by \cite{johansen15}. Models suggest that asteroids with diameters $D \gtrsim 120 \ {\rm km}$ are primordial---in the sense of not having suffered late-time collisional evolution \citep{bottke05}. The slope of the size distribution of these bodies is $q \simeq 4.5$, which is significantly steeper than the streaming prediction. Smaller bodies are typically collisionally evolved. Modeling by \cite{tsirvoulis18} backs out a primordial size distribution for $D \lesssim 70 \ {\rm km}$ that is characterized by a slope $q = 2.43^{+0.07}_{-0.05}$, which is somewhat flatter than the simple power-law fit to streaming simulations but somewhat steeper than the low mass limit of the truncated power-law derived in this work. 

\subsection{Kuiper Belt size distribution}
\begin{figure}[t]
\vspace{0.0truein}
\begin{center}
\hspace*{-0.5cm}\includegraphics[width=0.52\textwidth,angle=0]{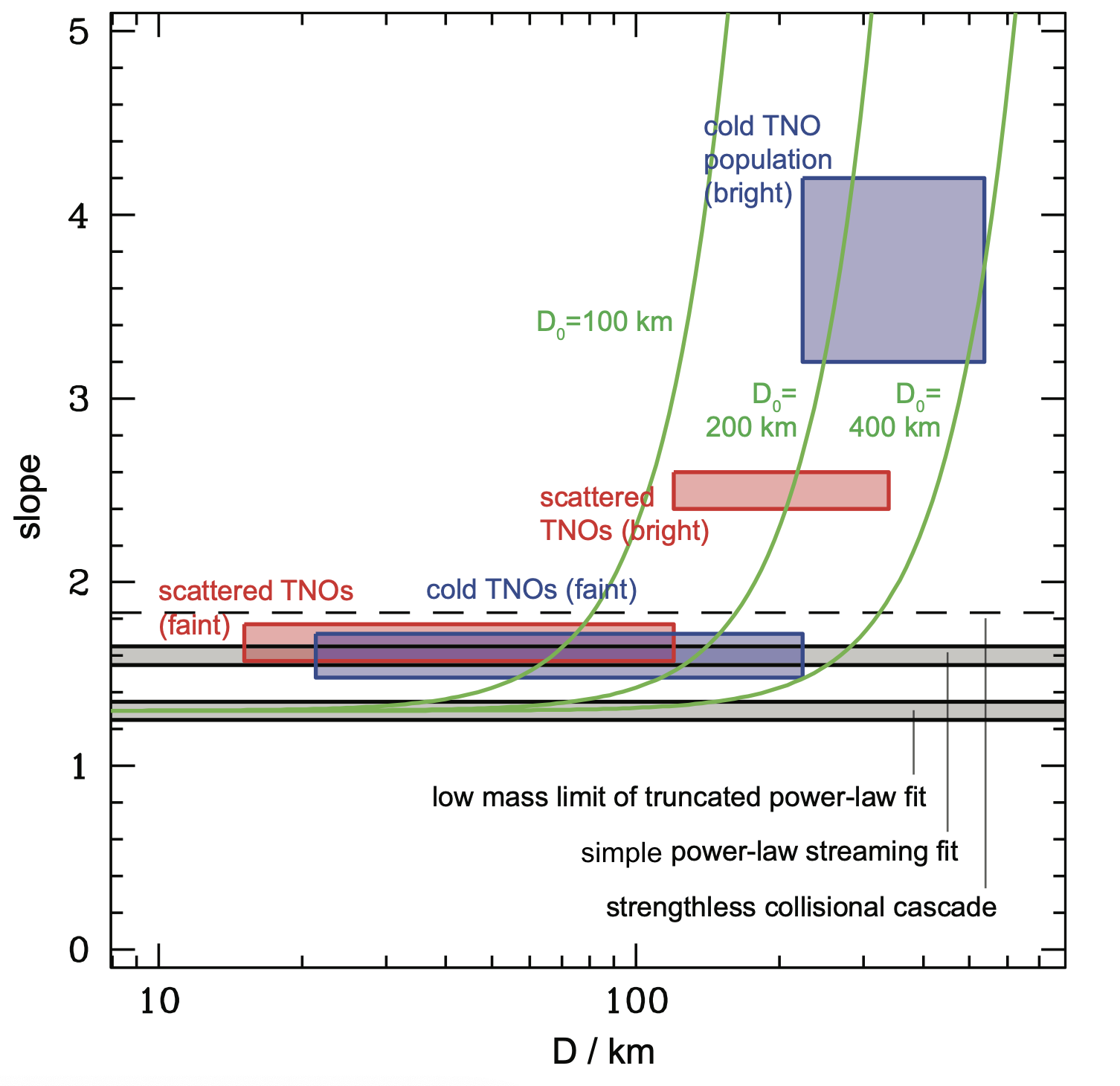}
\end{center}
\vspace{-0.2truein}
\caption{Comparison of theoretical estimates for the power-law index $p$ or $p^\prime$ of the initial planetesimal mass function with observational determinations for populations of Kuiper Belt Objects. The blue boxes show estimates for a dynamically cold population from 
\cite{fraser14}, assuming that the inferred break in the slope at $H_{\rm break} \simeq 6.9$ corresponds to $D \simeq 225 \ {\rm km}$. The red boxes (with arbitrarily assigned errors of $\Delta p = \pm 0.1$) show estimates for a population of scattered TNOs from \cite{lawler18}, assuming the inferred break in the slope at $H_{\rm break} \simeq 8.3$ corresponds to $D \simeq 120 \ {\rm km}$. The grey bands shows the best-fitting simple power-law fit ($p = 1.6 \pm 0.05$) and best-fitting low mass limit of the truncated power-law fit ($p^\prime = 1.3 \pm 0.05$) derived from this work. The green curves show the running of $p^\prime$ for the truncated power-law assuming a break radius of $D = 100$, 200, or 400~km. The dashed line shows the classical prediction for a strengthless collisional cascade \citep{dohnanyi69}.}
\label{KBOs}
\end{figure}

Turning to the Kuiper Belt, Figure~\ref{KBOs} shows estimates of $p$ derived from selected recent observational determinations of TNO size distributions. The blue boxes show estimates for a dynamically cold populations of TNOs from \cite{fraser14}. This population is arguably of the greatest interest theoretically, as modeling suggests that its properties are consistent with in situ formation with a low initial mass \citep{gomes18}. For their sample of cold TNOs \citep{fraser14} derive a faint-end slope $\alpha = 0.38^{+0.05}_{-0.09}$ ($p=1.63^{+0.09}_{-0.15}$), and a bright-end slope $\alpha = 1.5^{+0.4}_{-0.2}$ that has large uncertainties but which is clearly much steeper ($p \approx 3.5$). The inferred break occurs at an absolute magnitude $H \simeq 6.9$, which corresponds to $D \simeq 225 \ {\rm km}$ for an albedo of 0.04. The figure also shows, as the red boxes, estimates for $p$ for scattered TNOs derived by \cite{lawler18} using data from the Outer Solar System Origins Survey. (Note that the true uncertainty in $p$ for this sample is not clear; we have assigned arbitrary errors of $\Delta p = \pm 0.1$ in Fig.~\ref{KBOs}.) They derive a faint-end slope of $\alpha \simeq 0.4$ that is similar to that found for the cold population, and a bright-end slope of $\alpha \simeq 0.9$ that is shallower than for dynamically cold TNOs. The break between these populations occurs at $H \simeq 8.3$, corresponding to a diameter of about 120~km.

From Figure~\ref{KBOs} it is clear that the size distribution of TNOs in the range $10-20 \ {\rm km} \lesssim D \lesssim 10^2 \ {\rm km}$ is an excellent match to the best-fitting simple power-law derived from our (and other groups) simulation results. The agreement with the low mass limiting slope of the truncated power-law mass function (equation~\ref{cumulative}) is markedly worse, and the theoretical and observational results are significantly discrepant according to the purely statistical error bars that we quote. It is much harder, of course, to estimate the magnitude of possible systematic errors, arising for example from the fact that we fit only a limited range of planetesimal masses. The largest and brightest objects, conversely, display a much steeper distribution that appears to differ between dynamically distinct populations. The consistency of the steeper slopes at larger sizes with streaming predictions depends sensitively on the precise form of the high-mass cutoff. Using the simple exponential form that we have adopted we would infer a characteristic diameter (at the cutoff mass $M_0$) of $D_0 \sim 200 {\rm km}$. This, however, leads to very few large KBOs. A distribution more consistent with observations would require either a softer cutoff function \citep[such as that adopted by][]{schafer17} or a distribution that is the superposition of several mass functions with differing values of $M_0$.


\section{Conclusions}
\label{conclusions}

We have studied the mass distribution of planetesimals that form from gravitational collapse in aerodynamically coupled mixtures of solids and gas within protoplanetary disks. Gravitational collapse is part of several scenarios for planetesimal formation \citep[including the classical picture of fragmentation of a thin dust layer;][]{safronov72,goldreich73}, but appears easiest to realize as a secondary phase of particle clumping resulting from the streaming instability \citep{youdin05}. Accordingly, we have focused on simulation domains and physical parameters (in particular relatively high solid-to-gas ratios) that capture the regime where the streaming instability generates strong particle clumping \citep{carrera15,yang17}. The numerical setup of the simulations closely follows our earlier {\sc Athena}-based work \citep{simon16b,simon17} with one important addition: the use of a new clump-finding routine ({\sc Plan}) to robustly identify bound objects in three-dimensions. We also use a maximum likelihood estimator to fit an exponentially truncated power-law model to the simulation data, consistently extending our earlier power-law only results.

Our results are based on a set of high resolution simulations ($512^3$ or $512^2 \times 1024$ gas zones, $1.35 \times 10^8$ solid particles) that vary the strength of the background gas pressure gradient. We consider a range of $\Pi = \eta v_K / c_s$ between 0.0375 and 0.1, which is representative of the $\Pi$ values in a protoplanetary disk under typical assumptions, and separately model the $\Pi = 0$ limit, which is of interest because it is not subject to streaming instability. Our primary results are as follows.  

\begin{itemize}
\item[(i)]
For $\Pi \neq 0$, the simulations produce a mass function of planetesimals that, when fit as a simple power-law, yields an index $p = 1.5$ -- $1.6$ to within $2\sigma$.
\item[(ii)]
The slope of the derived mass functions clearly varies with mass, making a simple power-law a poor representation of the data. An exponentially truncated power-law provides a substantially better fit to the cumulative mass distribution. Using a maximum likelihood estimator we fit a function of the form $N(> M_p) \propto M_p^{-p^\prime+1} \exp[-M_p / M_0]$. The derived power-law index is generally consistent with $\pp \simeq 1.3$ across the $\Pi \neq 0$ runs, with a few exceptions.
\item[(iii)]
The exponentially truncated power-law model has a simple characteristic mass parameter $M_0$.
Characterizing the masses of the most massive planetesimals formed in the simulations using $M_0$, we find that $M_0$ is of the order of the gravitational mass $M_G$, which scales with disk parameters as $M_G \propto \Sigma_p^3 / \Omega^4$ (equation~\ref{grav_mass}). $M_0 / M_G$ increases linearly with $\Pi$, though zero dependence cannot be completely ruled out. Furthermore, $M_0/M_G$ may depend on whether self-gravity is operative from the start of the simulations, or instead turned on at a (mostly arbitrary) later time after non-linear clustering has already developed.
\item[(iv)]
 As noted by \cite{taki16} and \cite{youdin05} the linear scale of the streaming instability is proportional to $\Pi$, defining a mass scale that is proportional to $\Pi^3$. Our results exclude a scaling as steep as $M_0 \propto \Pi^3$. This suggests that the linear physics of the unstratified model problem does not simply translate into a prediction for the characteristic masses of planetesimals.
\item[(v)]
When combined with the results of prior simulations that varied the dimensionless stopping time $\tau$, the particle concentration $Z$ \citep{simon17}, and the strength of gravity relative to shear $\tilde{G}$ \citep{simon16a}, our results support the hypothesis that the initial mass function resulting from the collapse of over-densities is weakly dependent on the physical parameters relevant to the streaming instability. 
\item[(vi)]
The special case of $\Pi=0$ yields a mass function that is likely different than the streaming unstable cases, though the evidence for this remains modest considering the spread in $\pp$ values. We caution that the numerical challenges of modeling the $\Pi=0$ limit are somewhat different from the $\Pi > 0$ cases, and that more work is needed to study $\Pi=0$ in detail. Nonetheless, the observed differences constitute initial evidence that the streaming instability produces a distinct mass function that is not shared by other mechanisms that result in gravitational collapse.

\end{itemize}
 
Our work is subject to both numerical and physical limitations. On the numerical side, solving Poisson's equation on a uniform grid halts collapse on the grid scale at densities that are much smaller than material densities. This severely restricts our ability to say anything about the possible fragmentation of collapsing clumps into binaries (see \citealt{nesvorny10}), or clusters of smaller planetesimals. Furthermore, the dependence (if any) of the mass function on the level of {\em intrinsic} disk turbulence remains to be fully explored \citep[though see][]{johansen07a,johansen11}. We also note that, for reasons of numerical expediency, most work to date has focused on regions of parameter space where the non-self-gravitating streaming instability produces over-densities that would collapse rapidly in the presence of self-gravity \citep[as defined by][]{carrera15,yang17}. Precisely because of the speed of gravitational collapse, we might instead expect that physical planetesimal formation occurs near the threshold values of the parameters that allow it \citep{armitage16}. There is no evidence that a ``slow" approach to planetesimal formation---for example as particles increase their $\tau$ via coagulation or as a disk increases $Z$ locally in a pressure trap---changes the mass function, but neither can such possibilities be excluded based on current results.

Future work will need to quantify the systematic uncertainties in the predicted mass functions that arise from the above limitations. This is needed, most obviously, in order to assess whether the streaming predictions are compatible with observations of small body populations in the Solar System. Adopting our best fitting model (the truncated power law) we predict a slope for faint-end KBOs that is somewhat too flat ($p^\prime \approx 1.3$ versus the  $p^\prime \approx 1.6$ that is inferred observationally), and a cutoff that is substantially too sharp\footnote{The more gradual cutoff proposed by \cite{schafer17} would ameliorate this problem substantially. We defer study of the precise nature of the cutoff, however, to future work.}. Exactly how $p^\prime$ varies with mass is clearly sensitive to whether the observed objects formed in a single burst, or in multiple episodes with differing $M_0$, and hence the significance of the latter discrepancy is dependent on unknown aspects of Solar System planetesimal formation history. However, given the robustness of the slope at small sizes, as we have found here, this slope is a stronger, and possibly ultimately decisive, test.

From these considerations, the most important implication of a streaming origin for planetesimals remains the prediction of a top-heavy initial mass function \citep[e.g.][]{johansen07a,johansen15,simon16a,simon17,schafer17}. Compared to the classical alternative of km or sub-km scale planetesimals, a top-heavy mass function implies a dynamically hotter population that is accreted by growing protoplanets and giant planet cores more slowly. For model builders the apparent scaling of the characteristic or maximum mass $M_0$ with $M_G$ (implying in turn a steep---nominally cubic---scaling with particle surface density) may also be important. The main challenge in developing streaming-based models for forming planetary systems is the requirement for at least modestly super-Solar solid-to-gas ratios \citep{johansen09c,carrera15,yang17}. This result has also been found in all work to date, and it suggests two basic scenarios for successful planetesimal formation. One possibility is that a phase of {\em rapid} radial drift leads to a particle pile-up \citep{youdin04}, with planetesimal formation occurring at an early time primarily interior to some critical radius. Alternatively, the conditions needed to trigger gravitational collapse might be obtained only locally, in over-densities associated with zonal flows \citep{johansen09a,simon12}, vortices \citep{barge95}, or ice lines \citep{stevenson88,ida16,schoonenberg17,drazkowska17}. In some variations of this second scenario planetesimal formation would be directly linked to the mechanisms that {\em slow} radial drift, and there would be a direct connection between large-scale structure in disks and where and when planetesimals form.

The case for the streaming instability playing a central role in planetesimal formation has often hinged, in part, on the perceived shortcomings of other models. Further advances in the fidelity of numerical simulation, together with work to understand how to combine the results with models of particle growth and collisional evolution, offer the exciting prospect of instead comparing streaming predictions directly against Solar System and exoplanet data.


\acknowledgements
{\em Author contributions}. CPA ran the simulations and analyzed the results. JBS initiated and directed the project, and contributed to the analysis. RL developed the {\sc PLAN} code. All authors contributed to the interpretation and write-up of the results. We thank Daniel Wik for his insight into statistics and interpretation of our results. CPA, JBS and PJA acknowledge support from NASA under awards NNX13AI58G, NNX16AB42G and 80NSSC18K0640. JBS acknowledges support from NASA under {\em Emerging Worlds} through grant 80NSSC18K0597. RL acknowledges support from NASA under grant NNX16AP53H, and ANY acknowledges support from NSF AAS grant 1616929 and NASA ATP grant NNX17AK59G. The numerical simulations and analyses were performed on {\sc Stampede 2} through XSEDE grant TG-AST120062.

\bibliographystyle{apj}
\bibliography{ref}

\begin{thebibliography}{93}
\expandafter\ifx\csname natexlab\endcsname\relax\def\natexlab#1{#1}\fi

\bibitem[{{Akaike}(1974)}]{akaike74}
{Akaike}, H. 1974, IEEE Transactions on Automatic Control, 19, 716

\bibitem[{{Andrews} {et~al.}(2016){Andrews}, {Wilner}, {Zhu}, {Birnstiel},
  {Carpenter}, {P{\'e}rez}, {Bai}, {{\"O}berg}, {Hughes}, {Isella}, \&
  {Ricci}}]{andrews16}
{Andrews}, S.~M., {Wilner}, D.~J., {Zhu}, Z., {et~al.} 2016, \apjl, 820, L40

\bibitem[{{Armitage} {et~al.}(2016){Armitage}, {Eisner}, \&
  {Simon}}]{armitage16}
{Armitage}, P.~J., {Eisner}, J.~A., \& {Simon}, J.~B. 2016, \apjl, 828, L2

\bibitem[{Bai \& Stone(2010{\natexlab{a}})}]{bai10c}
Bai, X.-N., \& Stone, J.~M. 2010{\natexlab{a}}, The Astrophysical Journal, 722,
  1437

\bibitem[{Bai \& Stone(2010{\natexlab{b}})}]{bai10a}
---. 2010{\natexlab{b}}, The Astrophysical Journal Supplement, 190, 297

\bibitem[{Bai \& Stone(2010{\natexlab{c}})}]{bai10b}
---. 2010{\natexlab{c}}, The Astrophysical Journal Letters, 722, L220

\bibitem[{{Bai} \& {Stone}(2014)}]{bai14}
{Bai}, X.-N., \& {Stone}, J.~M. 2014, \apj, 796, 31

\bibitem[{{Barge} \& {Sommeria}(1995)}]{barge95}
{Barge}, P., \& {Sommeria}, J. 1995, \aap, 295, L1

\bibitem[{{Barnes} \& {Hut}(1986)}]{barnes86}
{Barnes}, J., \& {Hut}, P. 1986, \nat, 324, 446

\bibitem[{{B{\'e}thune} {et~al.}(2017){B{\'e}thune}, {Lesur}, \&
  {Ferreira}}]{bethune17}
{B{\'e}thune}, W., {Lesur}, G., \& {Ferreira}, J. 2017, \aap, 600, A75

\bibitem[{Birnstiel {et~al.}(2012)Birnstiel, Klahr, \& Ercolano}]{birnstiel12}
Birnstiel, T., Klahr, H., \& Ercolano, B. 2012, Astronomy and Astrophysics,
  539, A148

\bibitem[{Birnstiel {et~al.}(2011)Birnstiel, Ormel, \& Dullemond}]{birnstiel11}
Birnstiel, T., Ormel, C.~W., \& Dullemond, C.~P. 2011, Astronomy and
  Astrophysics, 525, 11

\bibitem[{{Blum}(2018)}]{blum18}
{Blum}, J. 2018, \ssr, 214, 52

\bibitem[{Blum \& Wurm(2008)}]{blum08}
Blum, J., \& Wurm, G. 2008, Annual Review of Astronomy and Astrophysics, 46, 21

\bibitem[{Bottke {et~al.}(2005)Bottke, Durda, Nesvorn{\'{y}}, Jedicke,
  Morbidelli, Vokrouhlick{\'{y}}, \& Levison}]{bottke05}
Bottke, W.~F., Durda, D.~D., Nesvorn{\'{y}}, D., {et~al.} 2005, Icarus, 175,
  111

\bibitem[{Carrera {et~al.}(2015)Carrera, Johansen, \& Davies}]{carrera15}
Carrera, D., Johansen, A., \& Davies, M.~B. 2015, arXiv.org

\bibitem[{Clauset {et~al.}(2009)Clauset, Shalizi, \& Newman}]{clauset09}
Clauset, A., Shalizi, C.~R., \& Newman, M. E.~J. 2009, SIAM Review

\bibitem[{Colella(1990)}]{colella90}
Colella, P. 1990, JCP, 87, 171

\bibitem[{Colella \& Woodward(1984)}]{colella84}
Colella, P., \& Woodward, P.~R. 1984, JCP, 54, 174

\bibitem[{{Cuzzi} {et~al.}(1993){Cuzzi}, {Dobrovolskis}, \&
  {Champney}}]{cuzzi93}
{Cuzzi}, J.~N., {Dobrovolskis}, A.~R., \& {Champney}, J.~M. 1993, \icarus, 106,
  102

\bibitem[{{Dittrich} {et~al.}(2013){Dittrich}, {Klahr}, \&
  {Johansen}}]{dittrich13}
{Dittrich}, K., {Klahr}, H., \& {Johansen}, A. 2013, \apj, 763, 117

\bibitem[{{Dohnanyi}(1969)}]{dohnanyi69}
{Dohnanyi}, J.~S. 1969, \jgr, 74, 2531

\bibitem[{{Dr{\c a}{\.z}kowska} \& {Alibert}(2017)}]{drazkowska17}
{Dr{\c a}{\.z}kowska}, J., \& {Alibert}, Y. 2017, Astronomy and Astrophysics,
  608, A92

\bibitem[{{Dubrulle} {et~al.}(1995){Dubrulle}, {Morfill}, \&
  {Sterzik}}]{dubrulle95}
{Dubrulle}, B., {Morfill}, G., \& {Sterzik}, M. 1995, \icarus, 114, 237

\bibitem[{{Eisenstein} \& {Hut}(1998)}]{eisenstein98}
{Eisenstein}, D.~J., \& {Hut}, P. 1998, \apj, 498, 137

\bibitem[{{Fortier} {et~al.}(2013){Fortier}, {Alibert}, {Carron}, {Benz}, \&
  {Dittkrist}}]{fortier13}
{Fortier}, A., {Alibert}, Y., {Carron}, F., {Benz}, W., \& {Dittkrist}, K.-M.
  2013, \aap, 549, A44

\bibitem[{Fraser {et~al.}(2014)Fraser, Brown, Morbidelli, Parker, \&
  Batygin}]{fraser14}
Fraser, W.~C., Brown, M.~E., Morbidelli, A., Parker, A., \& Batygin, K. 2014,
  The Astrophysical Journal, 782, 100

\bibitem[{Gardiner \& Stone(2005)}]{gardiner05a}
Gardiner, T.~A., \& Stone, J.~M. 2005, JCP, 205, 509

\bibitem[{Gardiner \& Stone(2008)}]{gardiner08}
---. 2008, JCP, 227, 4123

\bibitem[{{Goldreich} \& {Ward}(1973)}]{goldreich73}
{Goldreich}, P., \& {Ward}, W.~R. 1973, \apj, 183, 1051

\bibitem[{{Gomes} {et~al.}(2018){Gomes}, {Nesvorn{\'y}}, {Morbidelli},
  {Deienno}, \& {Nogueira}}]{gomes18}
{Gomes}, R., {Nesvorn{\'y}}, D., {Morbidelli}, A., {Deienno}, R., \&
  {Nogueira}, E. 2018, \icarus, 306, 319

\bibitem[{{Gundlach} \& {Blum}(2015)}]{gundlach15}
{Gundlach}, B., \& {Blum}, J. 2015, \apj, 798, 34

\bibitem[{{G{\"u}ttler} {et~al.}(2010){G{\"u}ttler}, {Blum}, {Zsom}, {Ormel},
  \& {Dullemond}}]{guttler10}
{G{\"u}ttler}, C., {Blum}, J., {Zsom}, A., {Ormel}, C.~W., \& {Dullemond},
  C.~P. 2010, \aap, 513, A56

\bibitem[{Hawley {et~al.}(1995)Hawley, Gammie, \& Balbus}]{hawley95a}
Hawley, J.~F., Gammie, C.~F., \& Balbus, S.~A. 1995, ApJ, 440, 742

\bibitem[{{Ida} \& {Guillot}(2016)}]{ida16}
{Ida}, S., \& {Guillot}, T. 2016, Astronomy and Astrophysics, 596, L3

\bibitem[{{Isella} {et~al.}(2016){Isella}, {Guidi}, {Testi}, {Liu}, {Li}, {Li},
  {Weaver}, {Boehler}, {Carperter}, {De Gregorio-Monsalvo}, {Manara}, {Natta},
  {P{\'e}rez}, {Ricci}, {Sargent}, {Tazzari}, \& {Turner}}]{isella16}
{Isella}, A., {Guidi}, G., {Testi}, L., {et~al.} 2016, Physical Review Letters,
  117, 251101

\bibitem[{{Johansen} {et~al.}(2011){Johansen}, {Klahr}, \&
  {Henning}}]{johansen11}
{Johansen}, A., {Klahr}, H., \& {Henning}, T. 2011, \aap, 529, A62

\bibitem[{Johansen {et~al.}(2015)Johansen, Mac~Low, Lacerda, \&
  Bizzarro}]{johansen15}
Johansen, A., Mac~Low, M.-M., Lacerda, P., \& Bizzarro, M. 2015, Science
  Advances, 1, 1500109

\bibitem[{Johansen {et~al.}(2007)Johansen, Oishi, Mac~Low, Klahr, Henning, \&
  Youdin}]{johansen07a}
Johansen, A., Oishi, J.~S., Mac~Low, M.-M., {et~al.} 2007, Nature, 448, 1022

\bibitem[{Johansen {et~al.}(2009{\natexlab{a}})Johansen, Youdin, \&
  Klahr}]{johansen09a}
Johansen, A., Youdin, A., \& Klahr, H. 2009{\natexlab{a}}, The Astrophysical
  Journal, 697, 1269

\bibitem[{Johansen {et~al.}(2009{\natexlab{b}})Johansen, Youdin, \&
  Mac~Low}]{johansen09c}
Johansen, A., Youdin, A., \& Mac~Low, M.-M. 2009{\natexlab{b}}, The
  Astrophysical Journal Letters, 704, L75

\bibitem[{Johansen {et~al.}(2012)Johansen, Youdin, \& Lithwick}]{johansen12}
Johansen, A., Youdin, A.~N., \& Lithwick, Y. 2012, Astronomy and Astrophysics,
  537, A125

\bibitem[{Johnson {et~al.}(2008)Johnson, Guan, \& Gammie}]{johnson08}
Johnson, B.~M., Guan, X., \& Gammie, C.~F. 2008, ApJS, 179, 553

\bibitem[{Kass \& Raftery(1995)}]{kass95}
Kass, R.~E., \& Raftery, A.~E. 1995, Journal of the American Statistical
  Association, 90, 773

\bibitem[{{Koyama} \& {Ostriker}(2009)}]{koyoma09}
{Koyama}, H., \& {Ostriker}, E.~C. 2009, \apj, 693, 1316

\bibitem[{{Krapp} {et~al.}(2019){Krapp}, {Ben{\'\i}tez-Llambay}, {Gressel}, \&
  {Pessah}}]{krapp19}
{Krapp}, L., {Ben{\'\i}tez-Llambay}, P., {Gressel}, O., \& {Pessah}, M.~E.
  2019, \apjl, 878, L30

\bibitem[{{Krivov} \& {Booth}(2018)}]{krivov18}
{Krivov}, A.~V., \& {Booth}, M. 2018, \mnras, 479, 3300

\bibitem[{{Lawler} {et~al.}(2018){Lawler}, {Shankman}, {Kavelaars},
  {Alexandersen}, {Bannister}, {Chen}, {Gladman}, {Fraser}, {Gwyn}, {Kaib},
  {Petit}, \& {Volk}}]{lawler18}
{Lawler}, S.~M., {Shankman}, C., {Kavelaars}, J.~J., {et~al.} 2018, \aj, 155,
  197

\bibitem[{{Li} {et~al.}(2019){Li}, {Youdin}, \& {Simon}}]{li19}
{Li}, R., {Youdin}, A., \& {Simon}, J. 2019, arXiv e-prints, arXiv:1906.09261

\bibitem[{{Li} {et~al.}(2018){Li}, {Youdin}, \& {Simon}}]{li18}
{Li}, R., {Youdin}, A.~N., \& {Simon}, J.~B. 2018, \apj, 862, 14

\bibitem[{Masset(2000)}]{masset00}
Masset, F. 2000, A\&AS, 141, 165

\bibitem[{Meerschaert {et~al.}(2012)Meerschaert, Roy, \& Shao}]{meerschaert12}
Meerschaert, M.~M., Roy, P., \& Shao, Q. 2012, Communications in Statistics -
  Theory and Methods, 41, 1839

\bibitem[{Morbidelli {et~al.}(2009)Morbidelli, Bottke, Nesvorn{\'{y}}, \&
  Levison}]{morbidelli09}
Morbidelli, A., Bottke, W.~F., Nesvorn{\'{y}}, D., \& Levison, H.~F. 2009,
  Icarus, 204, 558

\bibitem[{{Moro-Mart{\'{\i}}n} {et~al.}(2009){Moro-Mart{\'{\i}}n}, {Turner}, \&
  {Loeb}}]{moromartin09}
{Moro-Mart{\'{\i}}n}, A., {Turner}, E.~L., \& {Loeb}, A. 2009, \apj, 704, 733

\bibitem[{{Nakagawa} {et~al.}(1986){Nakagawa}, {Sekiya}, \&
  {Hayashi}}]{nakagawa86}
{Nakagawa}, Y., {Sekiya}, M., \& {Hayashi}, C. 1986, \icarus, 67, 375

\bibitem[{{Nesvorn{\'y}} {et~al.}(2010){Nesvorn{\'y}}, {Youdin}, \&
  {Richardson}}]{nesvorny10}
{Nesvorn{\'y}}, D., {Youdin}, A.~N., \& {Richardson}, D.~C. 2010, \aj, 140, 785

\bibitem[{Ormel \& Cuzzi(2007)}]{ormel07}
Ormel, C.~W., \& Cuzzi, J.~N. 2007, Astronomy and Astrophysics, 466, 413

\bibitem[{{Ormel} \& {Klahr}(2010)}]{ormel10}
{Ormel}, C.~W., \& {Klahr}, H.~H. 2010, \aap, 520, A43

\bibitem[{Partnership {et~al.}(2015)Partnership, Brogan, Perez, Hunter, Dent,
  Hales, Hills, Corder, Fomalont, Vlahakis, Asaki, Barkats, Hirota, Hodge,
  Impellizzeri, Kneissl, Liuzzo, Lucas, Marcelino, Matsushita, Nakanishi,
  Phillips, Richards, Toledo, Aladro, Broguiere, Cortes, {Cortes, P. C.},
  Espada, Galarza, Garcia-Appadoo, Guzman-Ramirez, Humphreys, Jung, Kameno,
  Laing, Leon, Marconi, Mignano, Nikolic, Nyman, Radiszcz, Remijan, Rodon,
  Sawada, Takahashi, Tilanus, Vila~Vilaro, Watson, Wiklind, Akiyama, Chapillon,
  de~Gregorio-Monsalvo, Di~Francesco, Gueth, Kawamura, Lee, Nguyen~Luong,
  Mangum, Pi{\'e}tu, Sanhueza, Saigo, Takakuwa, Ubach, Van~Kempen, Wootten,
  Castro-Carrizo, Francke, Gallardo, Garcia, Gonzalez, Hill, Kaminski, Kurono,
  Liu, L{\'o}pez, Morales, Plarre, Schieven, Testi, Videla, Villard, Andreani,
  Hibbard, \& Tatematsu}]{alma15}
Partnership, A., Brogan, C.~L., Perez, L.~M., {et~al.} 2015, arXiv.org, 2649

\bibitem[{Pinilla {et~al.}(2012)Pinilla, Birnstiel, Ricci, Dullemond, Uribe,
  Testi, \& Natta}]{pinilla12}
Pinilla, P., Birnstiel, T., Ricci, L., {et~al.} 2012, Astronomy and
  Astrophysics, 538, 114

\bibitem[{{Pollack} {et~al.}(1996){Pollack}, {Hubickyj}, {Bodenheimer},
  {Lissauer}, {Podolak}, \& {Greenzweig}}]{pollack96}
{Pollack}, J.~B., {Hubickyj}, O., {Bodenheimer}, P., {et~al.} 1996, \icarus,
  124, 62

\bibitem[{{Raymond} {et~al.}(2018){Raymond}, {Armitage}, {Veras}, {Quintana},
  \& {Barclay}}]{raymond18}
{Raymond}, S.~N., {Armitage}, P.~J., {Veras}, D., {Quintana}, E.~V., \&
  {Barclay}, T. 2018, \mnras, 476, 3031

\bibitem[{{Safronov}(1972)}]{safronov72}
{Safronov}, V.~S. 1972, {Evolution of the protoplanetary cloud and formation of
  the earth and planets.}

\bibitem[{Sch{\"a}fer {et~al.}(2017)Sch{\"a}fer, Yang, \& Johansen}]{schafer17}
Sch{\"a}fer, U., Yang, C.-C., \& Johansen, A. 2017, Astronomy and Astrophysics,
  597, A69

\bibitem[{{Schaffer} {et~al.}(2018){Schaffer}, {Yang}, \&
  {Johansen}}]{schaffer18}
{Schaffer}, N., {Yang}, C.-C., \& {Johansen}, A. 2018, \aap, 618, A75

\bibitem[{{Schoonenberg} \& {Ormel}(2017)}]{schoonenberg17}
{Schoonenberg}, D., \& {Ormel}, C.~W. 2017, Astronomy and Astrophysics, 602,
  A21

\bibitem[{{Sekiya} \& {Onishi}(2018)}]{sekiya18}
{Sekiya}, M., \& {Onishi}, I.~K. 2018, \apj, 860, 140

\bibitem[{Simon(2016)}]{simon16b}
Simon, J.~B. 2016, The Astrophysical Journal Letters, 827, L37

\bibitem[{Simon \& Armitage(2014)}]{simon14}
Simon, J.~B., \& Armitage, P.~J. 2014, The Astrophysical Journal, 784, 15

\bibitem[{Simon {et~al.}(2016)Simon, Armitage, Li, \& Youdin}]{simon16a}
Simon, J.~B., Armitage, P.~J., Li, R., \& Youdin, A.~N. 2016, The Astrophysical
  Journal, 822, 55

\bibitem[{Simon {et~al.}(2017)Simon, Armitage, Youdin, \& Li}]{simon17}
Simon, J.~B., Armitage, P.~J., Youdin, A.~N., \& Li, R. 2017, The Astrophysical
  Journal Letters, 847, L12

\bibitem[{Simon {et~al.}(2012)Simon, Beckwith, \& Armitage}]{simon12}
Simon, J.~B., Beckwith, K., \& Armitage, P.~J. 2012, Monthly Notices of the
  Royal Astronomical Society, 422, 2685

\bibitem[{Simon {et~al.}(2011)Simon, Hawley, \& Beckwith}]{simon11a}
Simon, J.~B., Hawley, J.~F., \& Beckwith, K. 2011, ApJ, 730, 94

\bibitem[{{Stevenson} \& {Lunine}(1988)}]{stevenson88}
{Stevenson}, D.~J., \& {Lunine}, J.~I. 1988, \icarus, 75, 146

\bibitem[{Stone \& Gardiner(2010)}]{stone10}
Stone, J.~M., \& Gardiner, T.~A. 2010, ApJS, 189, 142

\bibitem[{Stone {et~al.}(2008)Stone, Gardiner, Teuben, Hawley, \&
  Simon}]{stone08}
Stone, J.~M., Gardiner, T.~A., Teuben, P., Hawley, J.~F., \& Simon, J.~B. 2008,
  The Astrophysical Journal Supplement, 178, 137

\bibitem[{{Takahashi} \& {Inutsuka}(2014)}]{takahashi14}
{Takahashi}, S.~Z., \& {Inutsuka}, S.-i. 2014, \apj, 794, 55

\bibitem[{Taki {et~al.}(2016)Taki, Fujimoto, \& Ida}]{taki16}
Taki, T., Fujimoto, M., \& Ida, S. 2016, Astronomy and Astrophysics, 591, A86

\bibitem[{{Trujillo} {et~al.}(2001){Trujillo}, {Jewitt}, \& {Luu}}]{trujillo01}
{Trujillo}, C.~A., {Jewitt}, D.~C., \& {Luu}, J.~X. 2001, \aj, 122, 457

\bibitem[{{Tsirvoulis} {et~al.}(2018){Tsirvoulis}, {Morbidelli}, {Delbo}, \&
  {Tsiganis}}]{tsirvoulis18}
{Tsirvoulis}, G., {Morbidelli}, A., {Delbo}, M., \& {Tsiganis}, K. 2018,
  \icarus, 304, 14

\bibitem[{{Umurhan} {et~al.}(2019){Umurhan}, {Estrada}, \& {Cuzzi}}]{umurhan19}
{Umurhan}, O.~M., {Estrada}, P.~R., \& {Cuzzi}, J.~N. 2019, arXiv e-prints,
  arXiv:1906.05371

\bibitem[{{Wada} {et~al.}(2009){Wada}, {Tanaka}, {Suyama}, {Kimura}, \&
  {Yamamoto}}]{wada09}
{Wada}, K., {Tanaka}, H., {Suyama}, T., {Kimura}, H., \& {Yamamoto}, T. 2009,
  \apj, 702, 1490

\bibitem[{{Ward}(1976)}]{ward76}
{Ward}, W.~R. 1976, in Frontiers of Astrophysics, ed. E.~H. {Avrett}, 1--40

\bibitem[{Weidenschilling(1977)}]{weidenschilling77b}
Weidenschilling, S.~J. 1977, MNRAS, 180, 57

\bibitem[{{Weidenschilling}(1995)}]{weidenschilling95}
{Weidenschilling}, S.~J. 1995, \icarus, 116, 433

\bibitem[{{Whipple}(1972)}]{whipple72}
{Whipple}, F.~L. 1972, in From Plasma to Planet, ed. A.~{Elvius}, 211

\bibitem[{{Yang} {et~al.}(2017){Yang}, {Johansen}, \& {Carrera}}]{yang17}
{Yang}, C.-C., {Johansen}, A., \& {Carrera}, D. 2017, \aap, 606, A80

\bibitem[{{Yang} {et~al.}(2018){Yang}, {Mac Low}, \& {Johansen}}]{yang18}
{Yang}, C.-C., {Mac Low}, M.-M., \& {Johansen}, A. 2018, \apj, 868, 27

\bibitem[{Youdin \& Johansen(2007)}]{youdin07a}
Youdin, A., \& Johansen, A. 2007, The Astrophysical Journal, 662, 613

\bibitem[{{Youdin}(2011)}]{youdin11}
{Youdin}, A.~N. 2011, \apj, 731, 99

\bibitem[{{Youdin} \& {Chiang}(2004)}]{youdin04}
{Youdin}, A.~N., \& {Chiang}, E.~I. 2004, \apj, 601, 1109

\bibitem[{Youdin \& Goodman(2005)}]{youdin05}
Youdin, A.~N., \& Goodman, J. 2005, The Astrophysical Journal, 620, 459

\bibitem[{Zsom {et~al.}(2010)Zsom, Ormel, G{\"u}ttler, Blum, \&
  Dullemond}]{zsom10}
Zsom, A., Ormel, C.~W., G{\"u}ttler, C., Blum, J., \& Dullemond, C.~P. 2010,
  Astronomy and Astrophysics, 513, A57

\end{thebibliography}

\end{document}